\documentclass[fleqn,usenatbib]{mnras}

\usepackage{newtxtext,newtxmath}
\usepackage{xcolor}
\usepackage[T1]{fontenc}

\DeclareRobustCommand{\VAN}[3]{#2}
\let\VANthebibliography\thebibliography
\def\thebibliography{\DeclareRobustCommand{\VAN}[3]{##3}\VANthebibliography}

\usepackage{graphicx}	% Including figure files
\usepackage{amsmath}	% Advanced maths commands
\usepackage{multirow}
\usepackage{adjustbox}
\usepackage{array}
\usepackage{newtxtext,newtxmath}
\usepackage{caption}
\usepackage{lscape}

\title[Information content of external streams]{Probing the dark matter haloes of external galaxies with stellar streams}

\author[M. Walder et al.]{Madison Walder,$^{1}$\thanks{m.walder@surrey.ac.uk}
Denis Erkal,$^{1}$
Michelle Collins,$^{1}$
David Mart{\'\i}nez-Delgado$^{2}$
\\
$^{1}$Department of Physics, University of Surrey, Guildford, GU2 7XH, UK\\
$^{2}$Instituto de Astrof{\'\i}sica de Andaluc{\'\i}a, CSIC, Glorieta de la Astronom{\'\i}a, 18080 Granada, Spain\\}

\date{Accepted XXX. Received YYY; in original form ZZZ}

\pubyear{2023}

\begin{document}
\label{firstpage}
\pagerange{\pageref{firstpage}--\pageref{lastpage}}
\maketitle

% Abstract of the paper
\begin{abstract}

\noindent Stellar streams have proven to be powerful tools for measuring the Milky Way’s gravitational potential and hence its dark matter halo.  In the coming years, Vera Rubin, Euclid, ARRAKIHS, and NGRST will uncover a plethora of streams around external galaxies. Although great in number, observations of these distant streams will often be limited to only the on-sky position of the stream. In this work, we explore how well we will be able to measure the dark matter haloes of these galaxies by fitting simplified mock streams with a variety of intrinsic and orbital properties in a range of data availability scenarios. We find that streams with multiple wraps around their host galaxy can constrain the overall radial profile and scale radius of the potential without radial velocities. In many other cases, a single radial velocity measurement often provides a significant boost to constraining power for the radial profile, scale radius, and enclosed mass of the dark matter halo. Given the wealth of data expected soon, this suggests that we will be able to measure the dark matter haloes of a statistically significant sample of galaxies with stellar streams in the coming years.

%This is a simple template for authors to write new MNRAS papers.The abstract should briefly describe the aims, methods, and main results of the paper.It should be a single paragraph not more than 250 words (200 words for Letters). No references should appear in the abstract.
\end{abstract}

% Select between one and six entries from the list of approved keywords.
% Don't make up new ones.

%%%%%%%%%%%%%%%%%%%%%%%%%%%%%%%%%%%%%%%%%%%%%%%%%%

%%%%%%%%%%%%%%%%% BODY OF PAPER %%%%%%%%%%%%%%%%%%

\section{Introduction}
%All papers should start with an Introduction section, which sets the workin context, cites relevant earlier studies in the field by \citet{Fournier1901},and describes the problem the authors aim to solve \citep[e.g.][]{vanDijk1902}.Multiple citations can be joined in a simple way like \citet{deLaguarde1903, delaGuarde1904}.
The properties of dark matter (DM) haloes are sensitive to the properties of the DM particle such as its mass and whether it self-interacts. In Lambda Cold Dark Matter ($\Lambda$CDM), galaxies are embedded in centrally dense, extended, triaxial dark matter haloes \citep{1997ApJ...490..493N}, which contain an abundance of subhaloes \citep[e.g.][]{1999ApJ...522...82K}.  For self-interacting dark matter (SIDM), the main haloes are less centrally dense and more spherical throughout \citep[e.g.][]{2018MNRAS.479..359S}. In warm dark matter (WDM), the main halo properties are similar to those of $\Lambda$CDM, but the density distribution is smoother and there are far fewer subhaloes \citep[e.g.][]{libeskind_dicintio_knebe_yepes_gottlöber_steinmetz_hoffman_martinez-vaquero_2013}. Similar to WDM, fuzzy DM can suppress the formation of small scale substructure and can also form soliton cores in the centers of haloes \citep[e.g.][]{2017PhRvD..95d3541H}. In order to distinguish between these different dark matter models we can study how they affect the visible baryonic substructures in a galaxy.

Stellar streams are well-suited for probing dark matter halo properties. They form when a satellite, such as a dwarf galaxy or a globular cluster, is tidally disrupted by the gravitational potential of its host galaxy.  The stars that were once bound to the satellite become unbound, and form a stream of stars \citep[e.g.][]{1999A&A...352..149C,2008MNRAS.387.1248K, 2012MNRAS.420.2700K}. Each stream roughly traces out an orbit in the host potential \citep[][]{Sanders_Binney2013} which makes them sensitive to the host's gravitational potential and hence its  \citep[e.g.][]{Johnston_1999}.  The variety of streams provide unique approaches to different aspects of dark matter.  Streams from an accreted dwarf galaxy are kinematically hotter and will wrap around their host, thus allowing us to probe DM halo shape and mass \citep{2014MNRAS.445L..31B}.  While globular cluster streams are kinematically cold and therefore perturbations can be spotted and studied, allowing us to probe DM subhalo populations \citep[e.g.][]{2002MNRAS.332..915I,2002ApJ...570..656J}.  

Nearly one hundred stellar streams have been observed around the Milky Way (MW) \citep[e.g.][]{2006ApJ...642L.137B,2006ApJ...643L..17G,2014MNRAS.442L..85K,2016MNRAS.463.1759B,2018MNRAS.481.3442M,2018ApJ...862..114S,2021ApJ...914..123I,2023MNRAS.520.5225M} thanks to many deep and wide sky surveys photometric surveys such as SDSS \citep{2000AJ....120.1579Y}, DES \citep{2016MNRAS.460.1270D}, spectroscopic surveys LAMOST \citep{2012RAA....12.1197C}, APOGEE \citep{2017AJ....154...94M}, and all-sky astrometry provided by the \textit{Gaia} mission \citep{2016A&A...595A...1G}, leading to accurate and precise dynamical 6D information for numerous MW streams. This abundance of high-quality data has allowed extensive studying and modelling of MW stellar streams, including for making constraints on its gravitational potential and detecting possible dark matter subhaloes  \citep[e.g.][etc.]{2013MNRAS.436.2386L, 2019MNRAS.486.2995M,  2019MNRAS.487.2685E, 2020ApJ...901...48N, 10.1093/mnras/stab210, 2022ApJ...926..107M}.  Fitting models to GD-1 and Palomar 5 stream data has revealed a nearly spherical inner halo for our galaxy \citep[e.g.][]{Koposov_2010,2015ApJ...803...80K, 2016ApJ...833...31B}.  While fits to the Orphan-Chenab and Sagittarius streams in the outer Milky Way suggest that the outer halo should be significantly flattened and misaligned with the Milky Way's disk \citep{2019MNRAS.487.2685E, 2021MNRAS.501.2279V, 2023MNRAS.521.4936K}.  The discrepancy between the inner and outer halo shapes of the MW motivates us to not only further test and constrain our DM models, but to also increase the number of galaxies we test them with.  

Therefore, we must extend our stream-fitting techniques to extragalactic streams. Using streams around other galaxies to measure their dark matter content would allow us to robustly test $\Lambda$CDM and other DM particle models with a statistical sample of galactic DM halo properties.  We have already observed and characterized numerous stellar streams around Andromeda due to its proximity \citep[][]{2009Natur.461...66M}.  With photometric and spectroscopic surveys such as PAndAS \citep{2018ApJ...868...55M} and SPLASH \citep{2009ApJ...705.1275G}, we have been able to obtain accurate radial velocity and distance measurements \citep[e.g.][]{10.1093/mnras/stw513, Preston_2019} for many of its streams.  This has allowed a number of constraints on M31's potential, making it a key reference point for stream modelling in external galaxies.
\cite{2013MNRAS.434.2779F} fit the giant stellar stream (GSS) in M31 with N-body simulations and existing GSS radial velocity and distance measurements with which they inferred the mass of M31's halo, $M_{200} \sim 10^{12.27 \pm 0.1} \rm M_\odot$. However, observing and modelling streams beyond the Local Group (LG) proves to be much more of a challenge.

In recent years, ground-based observations have reached the depth and coverage needed to probe low surface brightness structures beyond the Local Group.  In particular, the Stellar Tidal Stream Survey \cite{MartinezDelgado_2010} revealed a handful of the extragalactic streams predicted around galaxies within the Local Volume \citep[e.g.][]{Johnston_2008, Cooper_2010}, establishing the first attempt to systematically detect external tidal features.  Now known as the Stellar Stream Legacy Survey, \cite{2023A&A...671A.141M} have uncovered more than 100 streams around other MW-like galaxies along with several other streams and debris discovered through other observational means \citep[][]{Gilhuly_2022,MiroCarretero_2023}.  However, as these efforts are pushing the limitations of our current observing power ($\mu_{\rm lim} \sim 29$ mag $\rm arcsec^{-2}$), the information provided by most of these observations is limited.

Streams at distances $\geq$ 4-6 Mpc are mainly traced as diffuse light structures, meaning we do not have the information provided by resolved stellar populations such as a robust measurement of the morphology or a precise stream track for most external streams observed to date.  This, in turn, makes them more difficult to fit with models since only the brightest parts of the stream can be seen, i.e. the stream's extent is not smoothly covered which makes it challenging to infer its accretion history and properties of the progenitor.   

Additionally, there is currently no facility with which to obtain direct radial velocity measurements of extragalactic stream segments.  Therefore, we must rely on the presence of a planetary nebula or globular cluster within the stream to obtain its kinematics \citep[e.g.][]{Foster_2014}.  Even in this case, this approach limits us to the use of streams which come from massive progenitors.  \cite{Toloba_2016} attempt to remedy this issue by combining spectroscopic techniques used for individual stars and diffuse light to obtain the kinematics for the stellar stream around NGC 444.  However, this technique requires narrow-band photometry and spectroscopic follow-up on 8-10m class telescopes.

Despite these observational limitations, using the current state-of-the-art stream modelling machinery developed for studying Milky Way streams allows us to infer at least some properties of an extragalactic stream, such as how or when it may have formed, as well as properties of the halo it resides in.  The first attempt to fit an external galaxy's potential with streams (beyond the Local Group) was by \cite{2015arXiv150403697A}. They attempted to constrain NGC 1097's halo mass and inner/outer potential profile with a spherical broken power law potential by fitting its stream with a model of the progenitor satellite galaxy shedding particles as it orbits. Currently, a similar modelling technique, known as the `particle spray' technique, has become widely adopted for its ability to reproduce observed stream characteristics with nearly the same accuracy of N-body simulations, without being as computationally expensive \citep[e.g.][]{2013MNRAS.434.2779F,2014MNRAS.445.3788G}.  For example, this technique is employed by \cite{2021MNRAS.506.5030M} in which they are able to reproduce the stream observed around M104 and determine a possible explanation for how the debris was formed. Similarly, \cite{2019ApJ...883L..32V} fit a `particle spray' model to the stream in NGC 5907 and was able to reproduce the observed features.        

More recently, \cite{Pearson_2022}, used the `particle spray' technique to model the Dw3 stream around Centaurus A to fit its dark matter halo. They fit the stream's morphology, making use of the known distance to Cen A and enclosed mass from globular cluster kinematics, to obtain a lower bound on the mass of Cen A's DM halo including just one radial velocity measurement from the stream. Furthermore, \cite{2023arXiv230317406N} explored the constraints to be made on the gravitational potential of external galaxies by connecting the curvature of a stream's track to the host potential's gravitational acceleration. Using mock observations of N-body simulations, they showed the flattening of the halo as well as disk-to-halo mass ratios can be recovered using only the on-sky track of the stream. They applied this to the stream in NGC 5907 and found it to have an oblate halo. 

In the coming decades, thousands of extragalactic streams in the local Universe are expected to be observed with the Nancy Grace Roman Space Telescope (NGRST) \citep{2013arXiv1305.5422S}, Rubin Observatory \citep{Ivezić_2019}, Euclid \citep{2016SPIE.9904E..0OR}, ARRAKIHS \citep[][]{ARRAKIHS_proposal}, and MOSAIC \citep{2020SPIE11447E..25S}. However, this wealth of data will be of lower signal-to-noise and dimensionality compared with the Milky Way, often limiting us to only the on-sky positions of the stream.  Therefore, we need to determine what we can learn about extragalactic dark matter haloes given the limitations of current and future observations, i.e. the information content of extragalactic streams. In particular, in this work, we focus on how well we can measure the radial profile of an external galaxy's dark matter halo with stellar streams. 

In this paper, we aim to determine how much information we require in addition to on-sky stream tracks to (1) determine the properties of a galaxy's DM halo and (2) uncover the information stored in the intrinsic properties of extragalactic streams. We approach this by approximating simplified mock streams as orbits in 3 different potentials from an external point of view. We then take mock observations of the on-sky stream track in different data availability scenarios and run an MCMC to fit orbit models to them, repeating the process for streams with varying lengths, inclination angles, apocenters, and eccentricities.

The paper is organized as follows: In Section 2, we present our approach to generating a set of mock stream-host potential observational scenarios and describe our method of fitting them.  In Section 3, we present the results of the stream fits and the constraints we can make on the DM halo in each potential. In Section 4 we discuss the implications of our results and conclude in Section 5.      

\section{Method}
In this section, we describe our approach to gauging the information content of extragalactic streams. In Section \ref{stream_gen}, we describe how our mock streams (orbits) are created.  We then describe how we take mock observations of each mock stream and fit them to constrain the properties of their respective dark matter potentials in Section \ref{mock_obs_fit}. Finally, we detail how we follow this procedure for different data availability scenarios for a set of streams varying in length, inclination angle, apocenter, and eccentricity in Section \ref{mockstreams}.

%Normally the next section describes the techniques the authors used.
%It is frequently split into subsections, such as Section~\ref{sec:maths} below.

\subsection{Mock stream generation} \label{stream_gen}

\begin{table}
\caption{The properties of our host potentials. The left column describes the dark matter-only potential (DM), consisting of just an NFW halo. The middle column describes the components of the galaxy-like potential which consists of an NFW halo (NFW), a Miyamoto-Nagai disk (MN), and a Hernquist bulge (HQ) with their associated masses, scale radii, and scale heights. The right column describes the power-law potential which is driven by circular velocity $v_c$, scale radius $r_0$, and $\gamma$.}
\label{tab:host_props}
\begin{adjustbox}{width=\columnwidth}
\begin{tabular}{llr}
\multicolumn{3}{c}{\textbf{Host Potential Properties}} \\ \hline
\multicolumn{1}{l}{$\Phi_{\rm DM}$} & \multicolumn{1}{l}{$\Phi_{\rm Galaxy}$} & $\Phi_{\rm PL}$ \\ \hline
\multicolumn{1}{l}{$M_{\rm NFW} = 8 \times 10^{11} \rm M_{\odot}$} & \multicolumn{1}{l}{$\Phi_{\rm DM}$} & $v_c = 220$ km/s \\
\multicolumn{1}{l}{$r_{s} = 16$ kpc} & \multicolumn{1}{l}{$M_{\rm MN} = 6.8 \times 10^{10} \rm M_{\odot}$} & $r_0 = 8$ kpc \\
\multicolumn{1}{l}{$c = 15.3$} & \multicolumn{1}{l}{$a_{\rm MN} = 3$ kpc} & $\gamma = 0.1$ \\
\multicolumn{1}{l}{} & \multicolumn{1}{l}{$b_{\rm MN} = 0.28$ kpc} &  \\
\multicolumn{1}{l}{} & \multicolumn{1}{l}{$M_{\rm HQ} = 0.5 \times 10^{10} \rm M_{\odot}$} & \multicolumn{1}{l}{} \\
\multicolumn{1}{l}{} & \multicolumn{1}{l}{$r_{\rm HQ} = 0.5$ kpc} & 
\end{tabular}%
\end{adjustbox}

\end{table}

For simplicity and computational efficiency, we approximate streams as orbits. Although streams do not exactly follow orbits \citep[e.g.][]{Sanders_Binney2013}, this approach is still frequently used for fitting streams in the Milky Way \citep[e.g.][]{Koposov_2010, Hendel2017, 2019MNRAS.486.2995M}. Furthermore, given the limited dimensionality of the observational data we expect to have in the near future, this simplistic approach is well-suited to rapidly explore the information content of external streams and explore a large range of setups.

We model the orbits using a leap-frog integrator which is part of the stream modeling code developed in \cite{2019MNRAS.487.2685E} although we emphasize that we only use orbits and not streams. Similar to \cite{2021MNRAS.506.5030M}, we simulate the orbit in coordinates centered on the host galaxy and we then make a mock observation of the stream assuming that the observer is looking along the z-axis. In this work we use the x and y coordinates of the stream as the stream observables, i.e. we effectively place the stream at infinity since we ignore the trigonometric effects associated with the stream being at a finite distance from the observer. 

For our mock streams, in each chosen gravitational potential, the progenitor is launched from apocenter (chosen as $y = 0$ for simplicity) and integrated backwards in time from the starting position and velocity in steps of 0.1 Myr until it performs half of a wrap around the host galaxy, i.e. its winding angle (which we call the phase angle $\theta$), reaches $-\pi$. We then integrate an orbit forward in time until it reaches a phase angle of $\theta \approx \pi$ again to avoid any overlap and get a full wrap of a stream.

\begin{figure*}
    \centering
    \includegraphics[width=\textwidth]{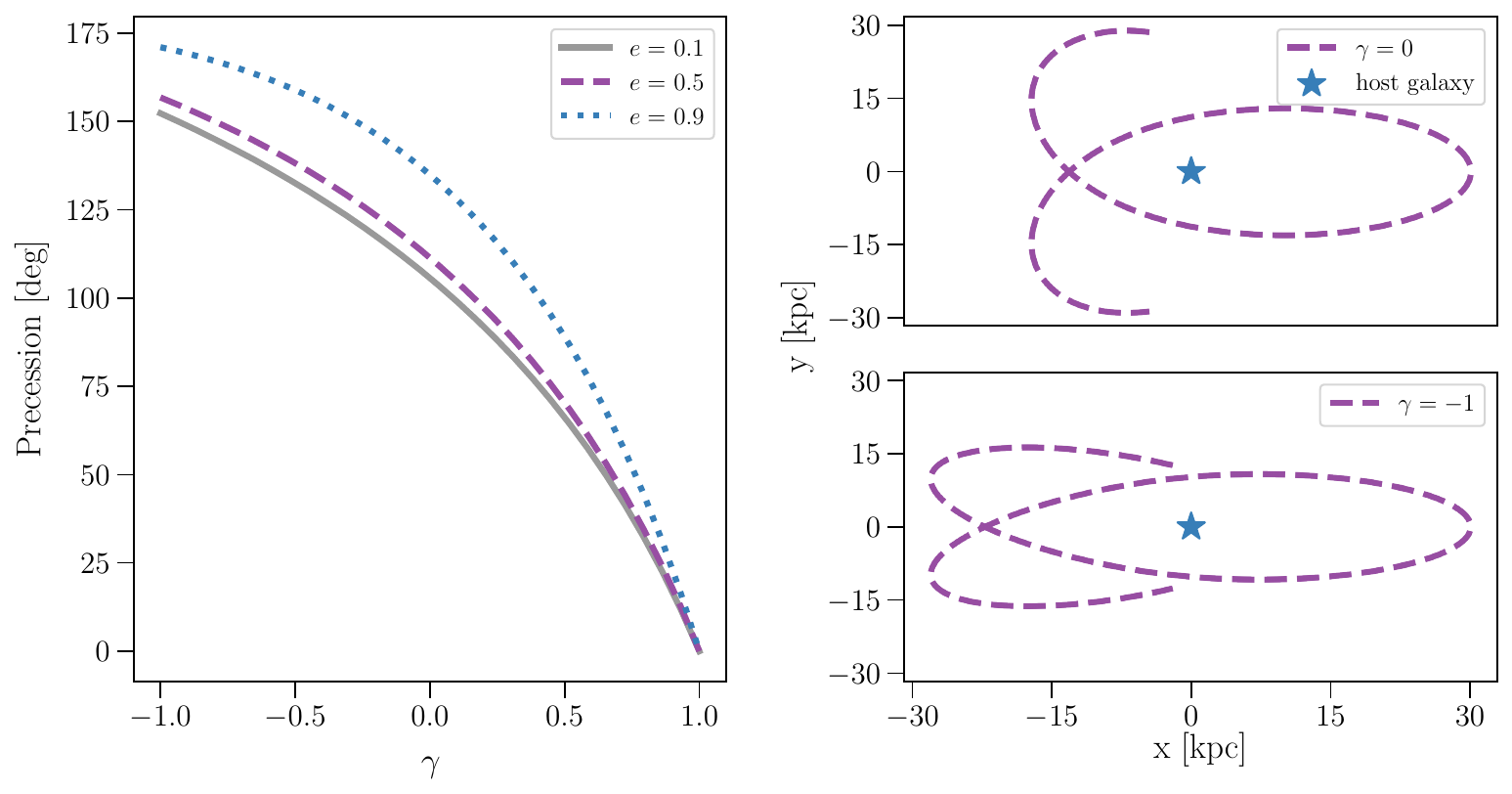}
    \caption{\textit{Left:}The angle of precession for orbits with same apocenter and different eccentricities as a function of $\gamma$ in $\Phi_{PL}$. \textit{Right:} a spatial representation of the same stream (with $e = 0.5$) in $\Phi_{PL}$ governed by $\gamma = 0$ in the top panel, and $\gamma = -1.$ in the bottom panel. Illustrates the relationship between a stream's properties and the gravitational potential it resides in. Motivated by Figure 9 (right panel) in \citet{2014MNRAS.437..116B}. }
    \label{fig:precession_vs_gamma}
\end{figure*}

Each orbit is integrated in the three gravitational potentials described in Table \ref{tab:host_props}. The $\Phi_{\rm DM}$ model is a dark matter-only model with an NFW profile \citep{1997ApJ...490..493N} whose parameters are similar to the \texttt{MWPotential2014} model in \cite{2015ApJS..216...29B}.  Next, $\Phi_{\rm Gal}$ is a more realistic galactic potential that contains dark matter and baryonic components similar to the Milky Way. It consists of a Miyamoto-Nagai disk \citep{1975PASJ...27..533M}, a Hernquist bulge \citep{1990ApJ...356..359H}, and an NFW halo and is nearly identical to \texttt{MWPotential2014} except we use a Hernquist bulge. Our final potential is a simplified power-law of the form:

\begin{equation} 
    \Phi_{\rm PL}(r) = -\frac{v_c^2}{\gamma}(\frac{r_0} {r})^{\gamma} ,
    \label{eq:pl_pot}
\end{equation}
where $r_0$ is a scale radius, $v_c$ is the circular velocity at the scale radius of the host, and $\gamma$ determines the radial form of the potential.  We use the first two potentials to evaluate how well $M_{\rm DM}$ and $r_{\rm s}$ can be recovered with and without the presence of baryons, while $\Phi_{\rm PL}$ is used to probe how well the overall radial profile of the potential can be recovered. Figure \ref{fig:precession_vs_gamma} illustrates the relationship between $\gamma$ and the precession of a stream's apocenter, showing how we are able to use the precession of the stream apocenters (i.e. apsidal precession) to measure the radial form of a potential since streams precess differently in different gravitational potentials \citep{2001ApJ...557..137J,2015MNRAS.454.2472H}. 

\subsection{Taking mock observations and fitting} \label{mock_obs_fit} 
\begin{figure*}
    \centering
    \includegraphics[width=\textwidth]{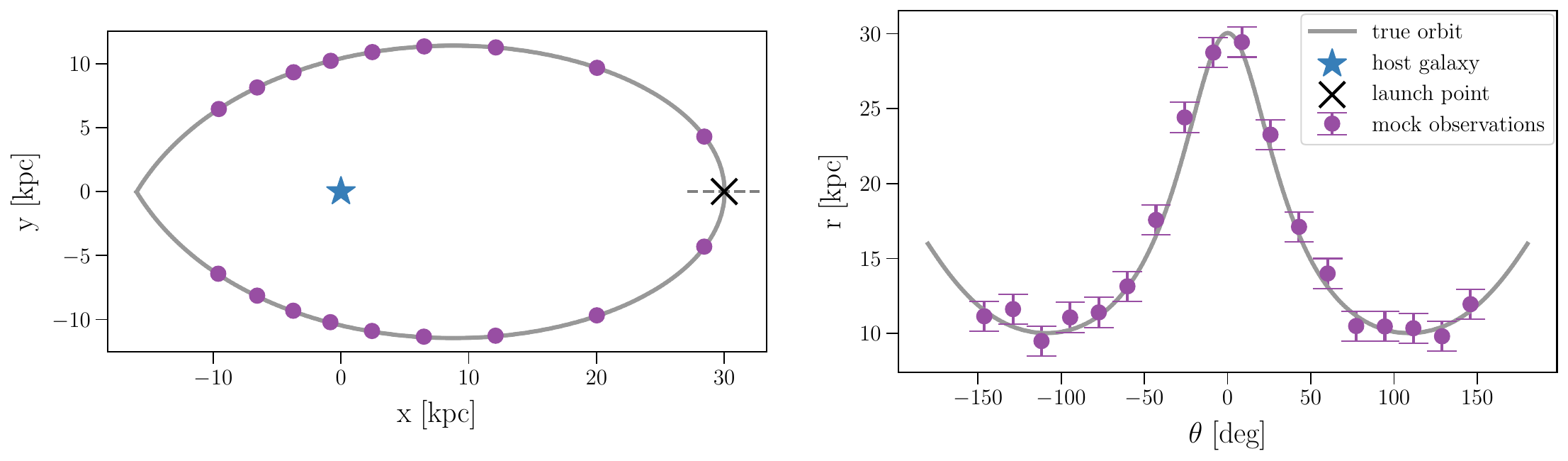}
    \caption{\textit{Left:} A spatial representation of the fiducial stream integrated in $\Phi_{\rm PL}$ (center marked by the blue star) as one would imagine observing it along the z-axis.  The purple points correspond to the points along the stream at which mock observations are taken.  The black cross with the small dashed line passing through indicates the approximate launch point for all of our stream models.  \textit{Right:} The fiducial stream in terms of 2D radius from the host galaxy in kpc and the position along the stream $\theta$ in radians.}
    \label{fig:mock_obs}
\end{figure*}

To create our mock observations, we transform each mock stream into projected `on-sky' coordinates, i.e. the 2-dimensional distance from the host ($r = \sqrt{x^2 + y^2}$), versus the angle along the stream $\theta$ which we visualize in Figure \ref{fig:mock_obs}.  We then take mock stream track observations along the orbit in segments of $\theta = 0.3$ rad ($\sim 17^{\circ}$) and measure the radius at each of these angles. We assume an observational error of $\sigma_{r} = 1$ kpc. This value is motivated by \cite{2021MNRAS.506.5030M} who found similarly sized errors when measuring the track of a stream in M104. We then repeat this procedure using an optimistic track uncertainty of $\sigma_{r} = 0.2$ kpc motivated by the requirement to resolve the widths of streams by future observational power in \cite{Laine_2018}. Additionally, we create a mock radial velocity measurement at $\theta=0$ (i.e. where the stream crosses the positive x-axis) assuming an observational uncertainty of $\sigma_{rv} = 10$ km/s. This uncertainty is motivated by studies of external streams using current ground-based telescopes \citep[e.g.][]{2013MNRAS.434.2779F, Escala_2020, Pearson_2022}. Note that since the line of sight direction is along the z-axis, our radial velocity is just $v_z$.

With these mock observables, we construct a likelihood function to compare them with a model stream:
\begin{equation} \label{orbit_likelihood}
    \mathscr{L} = \prod_{i = 1}^N\frac{1}{\sqrt{2\pi\sigma_{r}^2}}\exp\Big(-\frac{(r^{\rm obs.}_i - r^{\rm model}_i)^2}{2\sigma_{r}^2}\Big),
\end{equation}
where $r^{\rm obs.}_i$ is the `observed' radius of the mock stream in the angular bin $\theta_i$,  $r^{\rm model}_i$ is the radius of the model in the angular bin $\theta_i$, and $\sigma_r$ is the track uncertainty. We note that $r^{\rm model}_i$ is calculated by interpolating the model to where the `observed' data lies in $\theta$.  When including a mock radial velocity observation in the fit, we combine Equation \ref{orbit_likelihood} with a likelihood of similar form for $rv$ and $\sigma_{rv}$ in the place of $r$ and $\sigma_{r}$.

When generating orbits to compare with the mock streams, we fix one of the on-sky coordinates of the stream progenitor from which we launch orbits \citep[as in][]{2021MNRAS.506.5030M}. This is done since orbits launched from locations along the stream will produce equivalent observables which would result in a degeneracy. For simplicity, we fix the y-coordinate to $y=0$.

To properly explore the likelihood volume in parameter space, we implement an MCMC using {\fontfamily{qcr}\selectfont zeus} \citep{2021MNRAS.508.3589K} to find the progenitor's parameters ($x, z, v_x, v_y, v_z;$) as well as the host potential parameters, $M_{\rm NFW}$ and $r_s$ in $\Phi_{\rm DM}$ and $\Phi_{\rm Galaxy}$; $v_c$ and $\gamma$ in $\Phi_{\rm PL}$, which create an orbit model that best fits the observations. 
We define the priors on our varied parameters in Table \ref{tab:priors}, most of which are uniform distributions with the exception of normal distributions for the velocities. We also employ the following restrictions: the total velocity of the progenitor must be less than 1000 km/s, $0 < v_c < 1000$ km/s in $\Phi_{\rm PL}$, and the maximum orbit integration time must be < 14 Gyr to ensure a realistic model is created. Each MCMC run is initialized with 14 walkers, as recommended by {\fontfamily{qcr}\selectfont zeus} documentation, 10,000 total steps, and 5,000 burn-in steps.  

\begin{table}
\caption{The defined prior probabilities for our varied parameters}
\label{tab:priors}
\begin{adjustbox}{width=\columnwidth}
\begin{tabular}{|l|l|}
\hline
Parameter                    & Prior                                            \\ \hline
$x$, $z$                         & Uniform: \{-500, 500\} kpc                       \\
$v_{x}$, $v_y$, $v_z$, $v_c$ & Normal: \{$\mu = 0$, $\sigma = 250$\} km/s       \\
$\gamma$                     & Uniform: \{-1, 1\}                               \\
$M_{\rm NFW}$                    & Uniform: \{0, 200\} ($\times 10^{10} \,{\rm M}_{\odot}$) \\
$r_{\rm s}$                    & Uniform: \{0, 100\} kpc \\
\hline
\end{tabular}
\end{adjustbox}
\end{table}

\subsection{Data availability and stream scenarios} \label{mockstreams}
Next, we assess how well the dark matter halo properties of each stream-potential pair are recovered within the limit of 2D or 3D stream data (i.e. without or with a radial velocity) with current and expected future observational power by performing our procedure described in Sections \ref{stream_gen} and \ref{mock_obs_fit}. We do this for several different data availability scenarios: 
\begin{itemize}
    \item Stream track information only with $\sigma_{\rm track} = 1$ kpc
    \item Stream track (same $\sigma_{\rm track}$) + 1 radial velocity with $\sigma_{rv} = 10$ km/s
\end{itemize}
Then for the streams run in the $\Phi_{\rm Gal}$ potential, we also make and fit mock observations with a higher precision track uncertainty:
\begin{itemize}
    \item Stream track information only with $\sigma_{\rm track} = 0.2$ kpc
    \item Stream track (same $\sigma_{\rm track}$) + 1 radial velocity with $\sigma_{rv} = 10$ km/s
\end{itemize}
We note that each radial velocity measurement is taken at the location where the stream is launched, i.e. at $y=0$.

\begin{table}
\caption{The defined stream characteristics for various observational scenarios.  The length $L$ of the stream is defined in wraps, i.e. the number times it loops around the host.  The inclination angle $\phi$ of the stream towards or away from the observer is defined in degrees.  The 2D distance $D$ between the stream and its host in kpc.  The eccentricity $e$ of the stream.}
\label{tab:stream_props}
\begin{adjustbox}{width=\columnwidth}
\centering
\begin{tabular}{cccc}
\multicolumn{4}{c}{\textbf{Stream Cases}} \\ \hline
\multicolumn{1}{c}{$L$ [wraps]} & \multicolumn{1}{c}{$\phi$ [deg]} & \multicolumn{1}{r}{$r_{\rm apo}$ [kpc]} & $e$ \\ \hline
\multicolumn{1}{c}{[0.5, 1, 2]} & \multicolumn{1}{c}{45} & \multicolumn{1}{c}{30} & 0.4 \\
\multicolumn{1}{c}{{1}} & \multicolumn{1}{c}{{[0, 45, 70]}} & \multicolumn{1}{c}{{30}} & 0.4 \\
\multicolumn{1}{c}{1} & \multicolumn{1}{c}{45} & \multicolumn{1}{c}{[15, 30, 60]} & 0.4 \\
\multicolumn{1}{c}{1} & \multicolumn{1}{c}{45} & \multicolumn{1}{c}{30} & [0.2, 0.4, 0.7]
\end{tabular} \quad
\end{adjustbox}
\end{table}

Finally, we determine which stream properties provide the most information on their host potentials by repeating our approach for streams with varied lengths, inclinations, apocenters, and eccentricities as described by Table \ref{tab:stream_props} in the different data availability scenarios.  Every variation is made with respect to the fiducial case which we define as the stream with the following properties: $L = 1$ wrap around the host, inclined at $\phi = 45^{\circ}$ along the line of sight, apocenter of $r_{\rm apo} = 30$ kpc, and eccentricity of $e = 0.4$ (repeated values in Table \ref{tab:stream_props}).  For example, we vary the stream length while keeping the inclination angle, apocenter, and eccentricity the same as the fiducial. 

\begin{figure*}
    \centering
    \includegraphics[width=0.8\textwidth]{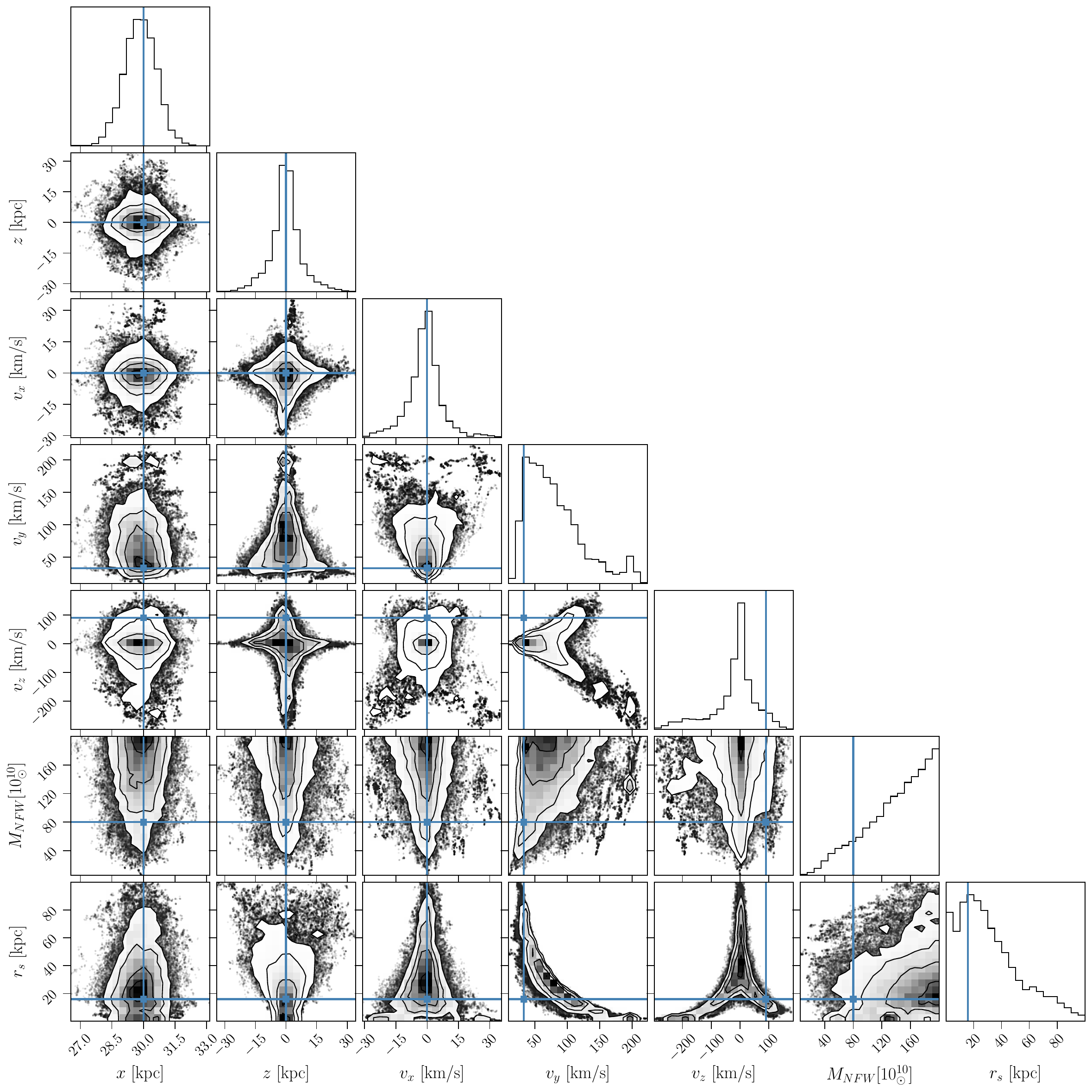}
    \caption{Corner plot of the MCMC samples for fitting the fiducial stream parameters $x, z, v_x, v_y, v_z,$ and $\Phi_{\rm DM}$ parameters $M_{\rm NFW}, r_s$ using only the mock observations of the stream track. The true values are indicated by the blue dots.  The posteriors of many of the fitted parameters are dominated by degeneracies.}
    \label{fig:cornerplot_PhiDM_example}
\end{figure*}

%Figures and tables should be placed at logical positions in the text. Don't
%worry about the exact layout, which will be handled by the publishers.

%Figures are referred to as e.g. Fig.~\ref{fig:example_figure}, and tables as
%e.g. Table~\ref{tab:example_table}.

\section{Results}
In this section, we present our results on the dark matter halo information obtained from mock observations of extragalactic streams with varying properties. In Section \ref{fiducialresults}, we detail our findings for the fiducial stream case for each of our test potentials and data availability scenarios.  In Section \ref{lengthresults}, we discuss what we learn when the stream length is changed, followed by the results for varying the inclination (Section \ref{incresults}), the distance of the stream's apocentre from the host (Section \ref{distresults}), and eccentricity (Section \ref{eccresults}).

\subsection{The Fiducial Stream} \label{fiducialresults}

For our fit to the fiducial stream in $\Phi_{\rm DM}$, neither the scale radius $r_{\rm s}$ nor the enclosed dark matter mass $M_{\rm DM}$ of the NFW profile can be recovered when fitting models using only the stream track with $\sigma_{\rm track} = 1$ kpc, as shown in the bottom right panels of Figure \ref{fig:cornerplot_PhiDM_example}.  This is largely due to the degeneracies encountered between halo and stream properties when working with stream data of limited dimensionality. The most notable of these degeneracies include those between halo mass and stream velocity and halo mass and scale radius. We will explore and discuss how they affect our results further in Section \ref{disc_degens_and_bias}.

Once a single radial velocity measurement is added to the fit, we can constrain the scale radius to be $r_s = 14.3^{+ 5.0}_{-4.3}$ kpc where the true value lies within one sigma.  In this case, the radial velocity measurement provides a large boost in constraining power, leading to an accurate and precise radial profile measurement.  Additionally, a radial velocity measurement improves the ability to recover $M_{\rm DM}$ with much greater precision.  However, we find our measured value of $M_{\rm DM}$ is $\sim 2$ times higher than the true value which lies just within 2~$\sigma$.  Some further investigation reveals this inflated mass measurement is due to a bias caused by how well the scale radius is constrained.  This bias affects a number of our results for the enclosed DM mass, which we will note throughout our findings as well as discuss further in Section \ref{disc_degens_and_bias}.

Similar to the case in $\Phi_{\rm DM}$, fitting only with the fiducial stream track in $\Phi_{\rm Gal}$ does not recover the scale radius of the DM halo nor the enclosed dark matter mass when the stream track uncertainty is 1 kpc.  This changes when a radial velocity measurement is added to the fit, allowing for a moderate scale radius constraint $r_s = 19.7^{+12.8}_{-7.7}$ kpc and a strong enclosed mass constraint $M_{\rm DM}[R = 30 \rm kpc] = 22.3^{+13.6}_{-10.0} \times 10^{10}\,{\rm M}_{\odot}$.  As our constraint on $r_s$ in this case is not incredibly precise, we believe the $M_{\rm DM}$ measurement is reflective of a true enclosed mass measurement with very little to no effect from the previously mentioned bias, which we explain further in Section \ref{disc_infocontent_rvs}.

Reducing the stream track to $\sigma_{\rm track} = 0.2$ kpc in $\Phi_{\rm Gal}$ allows the scale radius to be recovered without any radial velocity information $r_s = 12.7^{+11.7}_{-6.3}$ kpc.  However, the corresponding enclosed mass measurement is biased to be $\sim 2.3$ times more massive than the true $M_{\rm DM}$ which lies just outside of one sigma.  With a radial velocity, the scale radius constraint strengthens to $r_s = 16.7^{+4.9}_{-4.0}$ kpc, while the enclosed mass measurement becomes less inflated ($\sim 1.8$ times greater than the true value) and much more precise. 

Finally, when fitting the fiducial stream in $\Phi_{\rm PL}$, $\gamma$ remains unconstrained whether using only observations of the stream track with $\sigma_{\rm track} = 1$ kpc, or when incorporating a radial velocity measurement.  In both cases, the $\gamma$ posterior is skewed towards negative values due to a large volume in parameter space allowed by our priors, as well as a degeneracy between $\gamma$ and inclination angle.  Both of these will be discussed in more detail in Section \ref{disc_degens_and_bias}.   

\subsection{Changing Stream Length} \label{lengthresults}
\begin{figure*}
    \centering
    \includegraphics[width=0.8\textwidth] {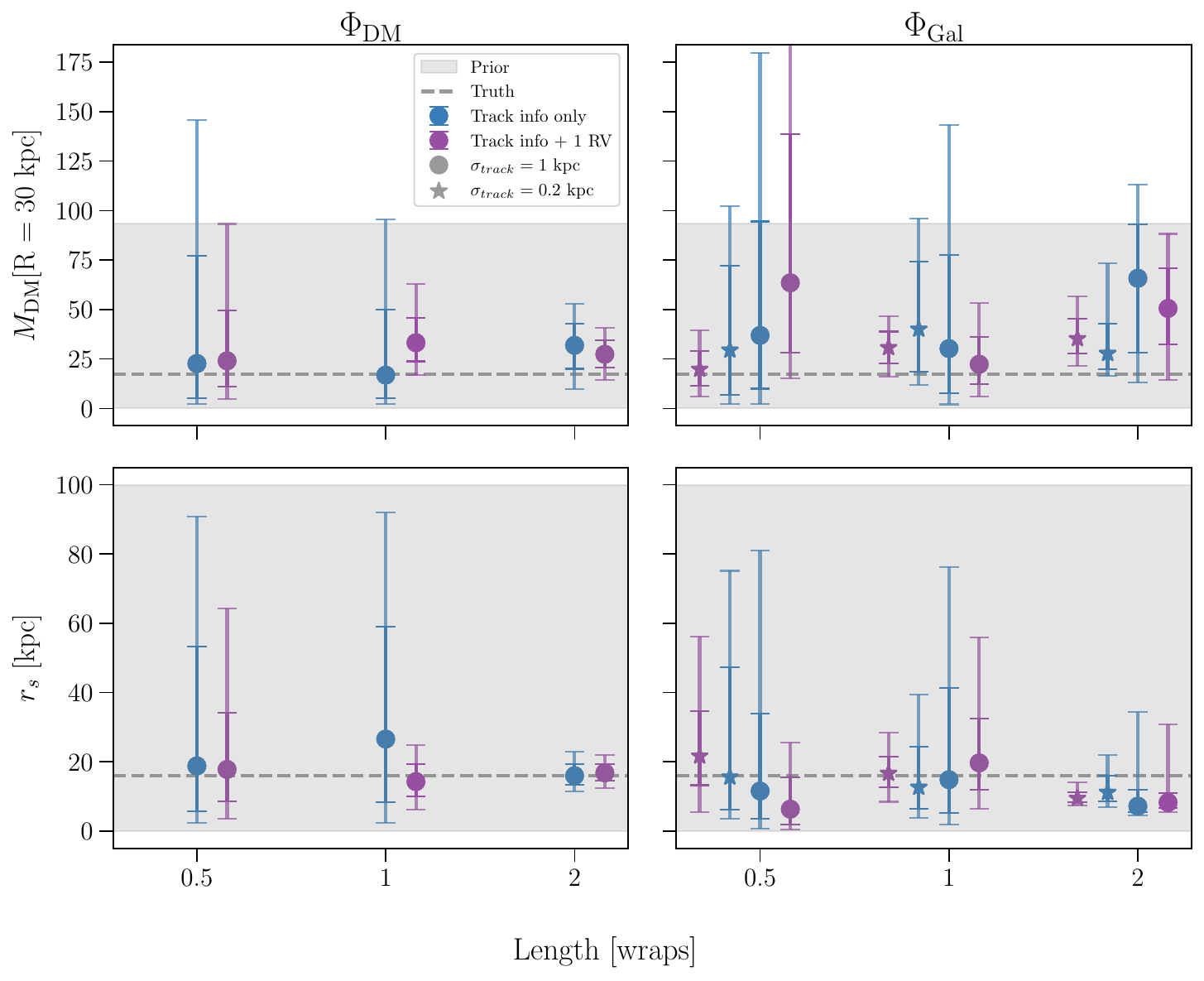}
    \caption{\textit{Top left:} The dependence of the enclosed dark matter mass measurement on the length of a stream for $\Phi_{\rm DM}$.  A gray, dashed line denotes the true value of $M_{\rm DM}$ at 30 kpc, while the colored circles and error bars correspond to the best fit $M_{\rm DM}$ value and its 1$\sigma$ and 2$\sigma$ uncertainties. The blue points correspond to fits made using only stream track measurements, while the purple ones denote track fits including a radial velocity measurement. \textit{Bottom left:} $r_s$ measurement as a function of stream length.  Similar to the panel above, the gray, dashed line shows the true value of $r_s$, while the colored points and associated uncertainties correspond to the best fit $r_s$ value.  The colors correspond to the same data availability scenarios as for the above panel as well as for the rest of the figure. \textit{Top right:} The relationship between enclosed mass and stream length for streams in $\Phi_{\rm Gal}$.  The circles correspond to the constraints made with a 1 kpc uncertainty in the stream track, while the star symbols corresponds to the constraints made with 0.2 kpc uncertainty in the stream track.  \textit{Bottom right:} The NFW scale radius measurement as a function of stream length for streams in $\Phi_{\rm Gal}$.}
    \label{fig:mass_and_rs_summary_varylength}
\end{figure*}

To assess the information stored in the length of an extragalactic stream, we vary the number of times a stream wraps around its host (first column of Table \ref{tab:stream_props}).  Figure \ref{fig:mass_and_rs_summary_varylength} shows the results of our fits to $M_{\rm DM}$ (top panels) and scale radius $r_s$ (bottom panels) of the dark matter halo for varying stream lengths in $\Phi_{\rm DM}$ (left side) and $\Phi_{\rm Gal}$ (right side). In $\Phi_{\rm DM}$, only a stream with two wraps around the host can provide a strong constraint on $r_s$ without radial velocity information ($r_s = 15.9^{+3.3}_{-2.5}$ kpc).  The measurement of $M_{\rm DM}$ for this case is also precise, but biased to be $\sim 1.8$x higher than the truth.  When a radial velocity is added to the fit, $r_s$ is constrained for all explored stream lengths with increasing precision as the length of the stream increases.  In particular, adding a radial velocity to the fit with the stream track with multiple loops slightly tightens the $r_s$ constraint, and lowers the $M_{\rm DM}$ measurement to be $\sim 1.6$x greater than $M_{\rm DM, true}$ with much smaller uncertainties. We argue a constraint on the range of mass enclosed can still be made in this case as both distributions are tight and precise with the true value just outside of 1$\sigma$.

In $\Phi_{\rm Gal}$, no constraints on the enclosed dark matter mass or scale radius are made for stream tracks observed with a $\sigma_{\rm track} = 1$ kpc uncertainty (circular points) without a radial velocity measurement included in the fit.  Adding a radial velocity allows for a tighter constraint on $r_s$ for the $L = 0.5$ wrap stream case. However, for a stream with multiple loops, the scale radius measurement is extremely precise but biased to lower values.  This is, again, due to the degeneracy between mass and velocity as we will explain further in Section \ref{disc_degens_and_bias}, resulting in biased enclosed mass measurements.  However, a lower bound on the dark matter mass can \textit{tentatively} be placed without a radial velocity, $M_{\rm DM}[R = 30 \rm kpc] \ge 8.5 \times 10^{10} M_{\odot}$ (discussed further in Section \ref{disc_infocontent_streamtrackonly}), and with a radial velocity the bound becomes $M_{\rm DM}[R = 30 \rm kpc] \ge 9.1 \times 10^{10} M_{\odot}$.

Smaller observational uncertainties ($\sigma_{\rm track} = 0.2$ kpc; star symbols in Fig. \ref{fig:mass_and_rs_summary_varylength}) allow for a moderate constraint on $r_s$ and a weak constraint on the enclosed mass when fitting the fiducial stream without a radial velocity measurement.  For a stream with 2 loops, the $r_s$ measurement and corresponding mass measurement is more precise than that measured with the larger stream track uncertainty, but not as biased.  When a radial velocity is added, $r_s$ can be constrained for all stream length cases with the exception of the stream with 2 wraps, with precision increasing as the stream gets longer.  The corresponding enclosed DM mass measurements are precise for every case when a radial velocity is included, but become inflated as the stream length gets longer and the $r_s$ constraints become more precise.       

\begin{figure}
    \centering
    \includegraphics[width=\columnwidth]{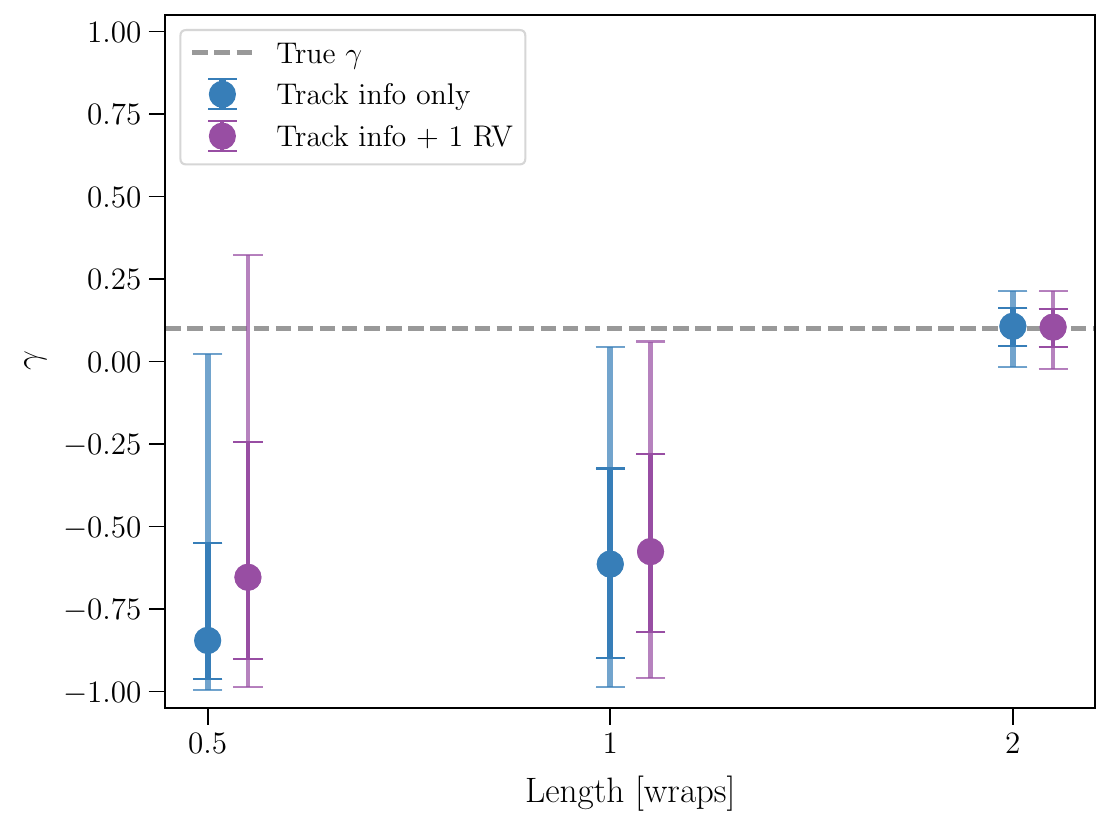}
    \caption{The constraint on the radial profile of a stream in $\Phi_{\rm PL}$  as a function of the length of the stream.  A gray, dashed line denotes the true value of $\gamma$, while the colored circles and error bars correspond to the best fit $\gamma$ value and its uncertainties out to 2$\sigma$. The blue points correspond to fits done using only stream track measurements, while the purple ones denote track fits including a radial velocity measurement.  Radial profile measurements improve with stream length, and $\gamma$ can be tightly constrained with and without a radial velocity.}
    \label{fig:gamma_summary_varylength}
\end{figure}

Figure \ref{fig:gamma_summary_varylength} displays the results of our fits to $\gamma$ in $\Phi_{PL}$ for each stream length, color-coded by data availability.  As mentioned in Sec. \ref{fiducialresults}, we often find our $\gamma$ measurements dominated by degeneracies.  Fortunately, we find the constraints on $\gamma$ improve as the streams get longer. Once the stream length reaches at least two wraps around the host, the degeneracies are completely overcome and we recover $\gamma = 0.1 \pm 0.06$ with and without a radial velocity measurement with nearly equal accuracy and precision.

\subsection{Changing Stream Inclination} \label{incresults}
\begin{figure*}
    \centering
    \includegraphics[width=0.8\textwidth] {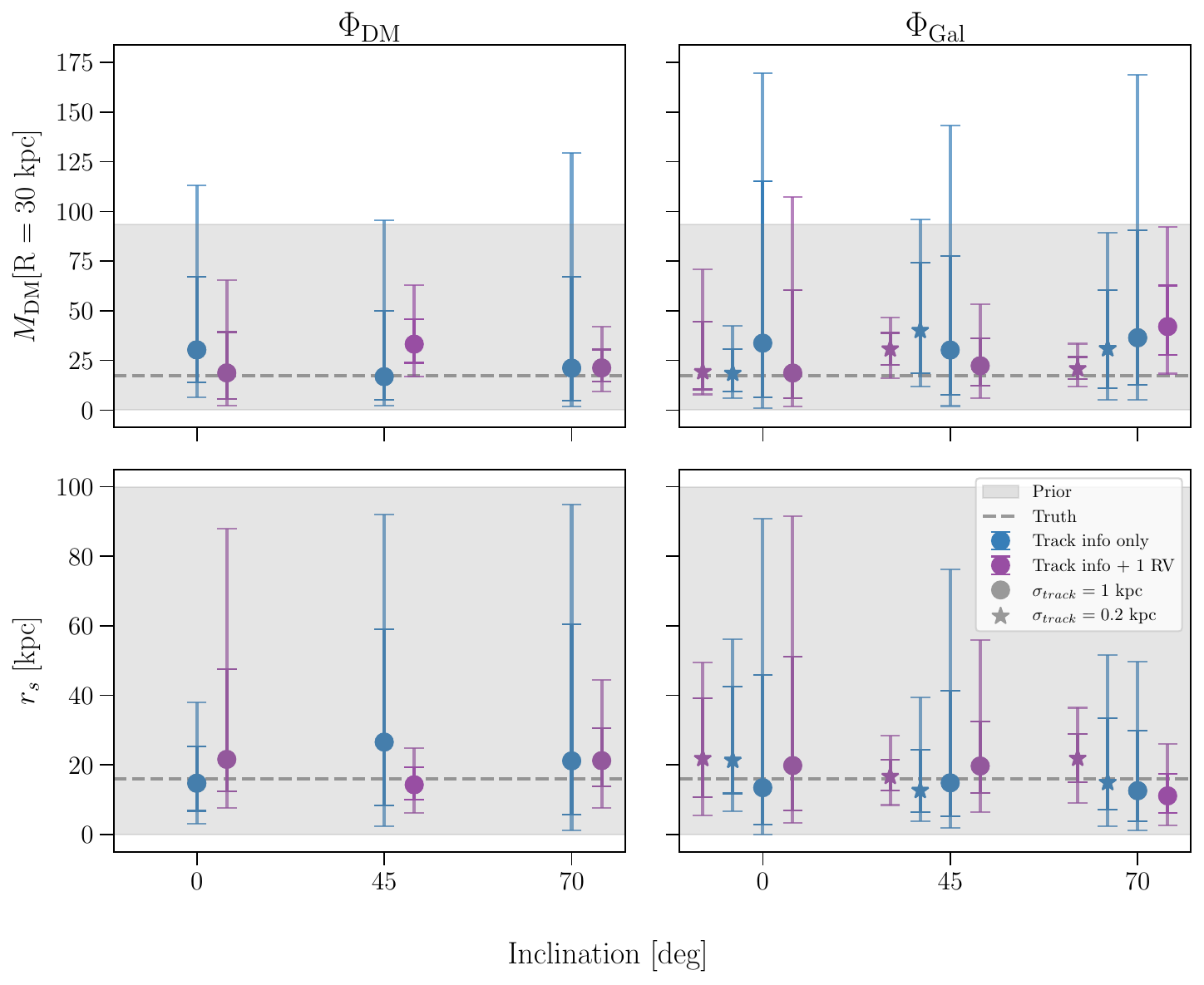}
    \caption{Same as Figure \ref{fig:mass_and_rs_summary_varylength} for varying stream inclination angles.}
    \label{fig:mass_and_rs_summary_varyinc}
\end{figure*}

We assess the information stored in the inclination angle of a stream by varying and fitting the fiducial orbit from face-on to nearly edge on (second column Table \ref{tab:stream_props}) viewing angles.  Figure \ref{fig:mass_and_rs_summary_varyinc} displays our results for enclosed DM mass and scale radius measurements as a function of inclination for both $\Phi_{\rm DM}$ and $\Phi_{\rm Gal}$, similar to Figure \ref{fig:mass_and_rs_summary_varylength}.
In $\Phi_{\rm DM}$, no constraints can be made on either $M_{\rm DM}$ or $r_s$ without radial velocity information, except for the face-on stream case in which $r_s$ is well-constrained.  Once a radial velocity is added, however, the constraint on the scale radius for this stream becomes much weaker while it strengthens for the more inclined streams. In addition, $M_{\rm DM}$ is weakly constrained for the face-on stream with a radial velocity but is measured more precisely as the inclination angle increases. The aforementioned bias between the $r_s$ constraint and $M_{\rm DM}$ does not seem to affect the face-on or most inclined orbit as much as it does the fiducial case since the $r_s$ measurements are less precise.  In particular, the measured enclosed dark matter mass for the $\phi = 70^{\circ}$ stream is $M_{\rm DM}[R = 30 \rm kpc] = 21.3^{+9.2}_{-6.7} \times 10^{10} \rm M_{\odot}$, having only a slight effect from the bias.

In $\Phi_{\rm Gal}$, there is a much more visible trend between $r_s$ and stream inclination as shown in the right side of Figure \ref{fig:mass_and_rs_summary_varyinc}.  For the fits made with $\sigma_{\rm track} = 1$ kpc, the scale radius can be recovered without any radial velocity information for the stream with $\phi = 70^{\circ}$, but the mass cannot.  The addition of a radial velocity to this case leads to a tighter $r_s$ constraint with a moderately biased $M_{\rm DM}$ measurement, and a lower bound can be placed at $M_{\rm DM}[R = 30 \rm kpc] \ge 8.5 \times 10^{10} \,{\rm M}_{\odot}$.

With smaller stream track uncertainties, weak constraints on $r_s$ can be made for both $\phi = 0^{\circ}, 70^{\circ}$ without radial velocities.  For the face-on stream, we can also make a strong constraint on the enclosed dark matter mass $M_{\rm DM}[R = 30 \rm kpc] = 18.5^{+12.3}_{-9.1} \times 10^{10} \rm M_{\odot}$.  Adding a radial velocity strengthens the $r_s$ constraint for the more inclined streams, as well as enables enclosed DM mass constraints for these cases with increasing precision as inclination increases (top right panel of Figure \ref{fig:mass_and_rs_summary_varyinc}).  To a lesser extent than the fiducial case, the $M_{\rm DM}$ measurement for the $\phi = 70^{\circ}$ stream case is only slightly affected by the bias as evinced by the tighter $r_s$ constraint, whereas $M_{\rm DM}$ for the face-on inclination is affected by it very little if at all.

\begin{figure}
    \centering
    \includegraphics[width=\columnwidth]{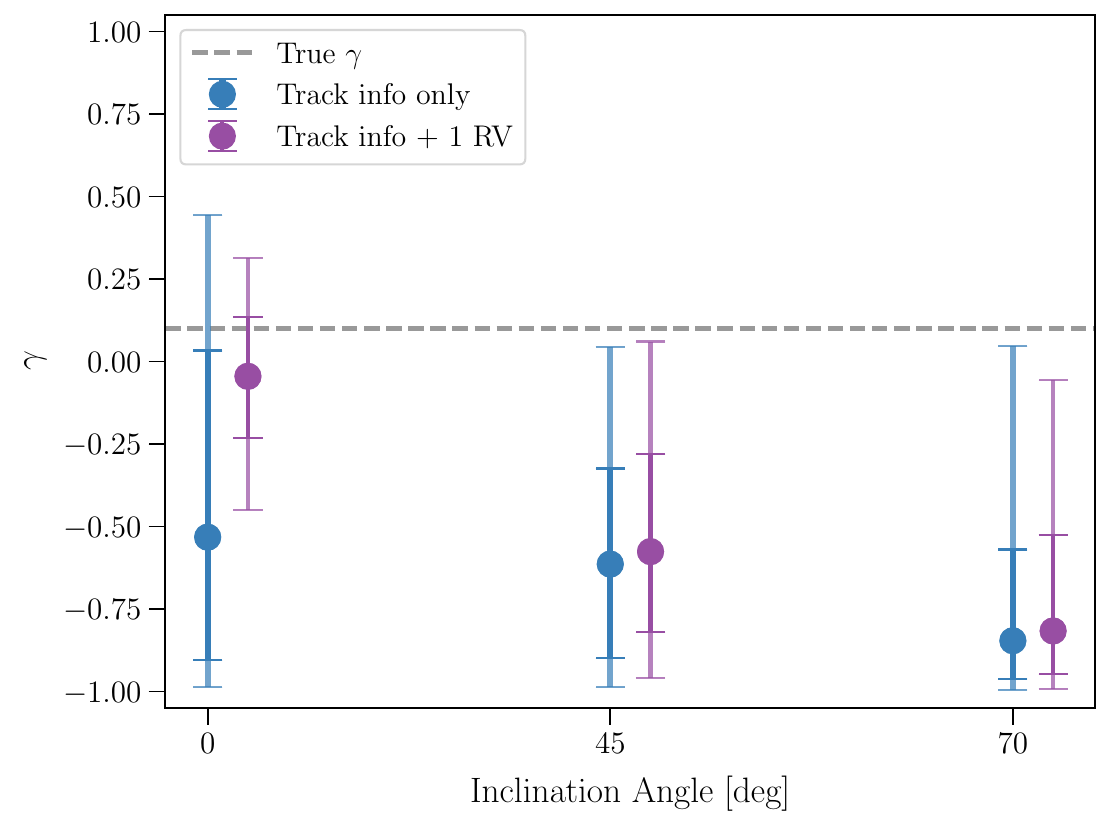}
    \caption{Similar to Fig. \ref{fig:gamma_summary_varylength}, the radial profile dependence on the inclination angle of an observed stream in $\Phi_{\rm PL}$.  Radial profile is best measured for observed streams with little inclination.}
    \label{fig:gamma_summary_varyinc}
\end{figure}

Figure \ref{fig:gamma_summary_varyinc} shows the results of our fits to constrain the radial profile $\gamma$ in $\Phi_{\rm PL}$ for these streams.  There is a clear trend between stream inclination angle and $\gamma$ in which $\gamma$ is better measured for streams at lower inclination. The measurements are generally dominated by the previously stated degeneracy between inclination angle and $\gamma$ until a radial velocity measurement is added to the orbit with $\phi = 0^{\circ}$, in which case $\gamma$ is measured to be $\gamma = 0 \pm 0.2$.

\subsection{Changing Stream Apocenter}
\label{distresults}
\begin{figure*}
    \centering
    \includegraphics[width=0.8\textwidth] {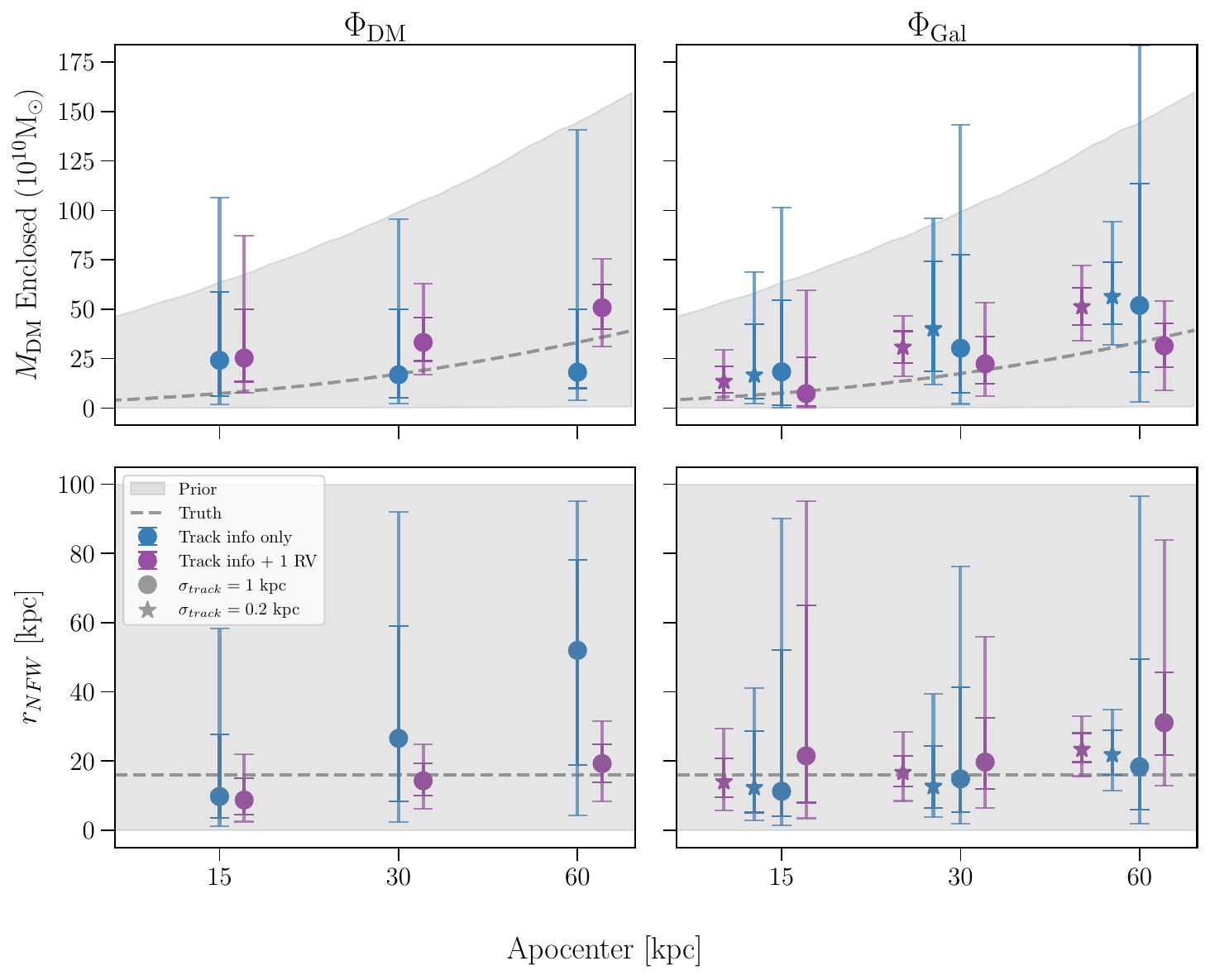}
    \caption{Similar to Figure \ref{fig:mass_and_rs_summary_varylength} for varying the distance of the stream's apocenter from the host galaxy. The key factor that distinguishes this figure from the others lies in the top two panels where the enclosed mass prior (gray shaded region) and true enclosed mass (dashed gray line) now vary as distance changes.}
    \label{fig:mass_and_rs_summary_varydist}
\end{figure*}

We gauge the information content contained in the distance of a stream to its host by varying the stream's apocenter from the fiducial case to be within the inner halo and then placed in the outer halo (third column Table \ref{tab:stream_props}).  Figure \ref{fig:mass_and_rs_summary_varydist} displays our $M_{\rm DM}$ and $r_s$ results for streams with varying apocenters.  Contrary to the previous figures, the top panels now show the enclosed dark matter mass as a function of distance from the host galaxy.  Without a radial velocity, only the scale radius for the stream with $r_{\rm apo} = 15$ kpc can be somewhat constrained in $\Phi_{\rm DM}$.  
 
With a radial velocity, $r_s$ can be recovered precisely at all explored apocenters with improved accuracy for streams with apocenters outside of the true scale radius.  Once a radial velocity is added, $M_{\rm DM}$ can be recovered at apocenters beyond the true scale radius, but they are each moderately biased by the high precision of the $r_s$ constraint. 

For the same streams in $\Phi_{\rm Gal}$ (right side of Fig. \ref{fig:mass_and_rs_summary_varydist}), no constraints can be made on $r_s$ or $M_{\rm DM}$ without a radial velocity measurement included in the fit for data with $\sigma_{\rm track} = 1$ kpc.  The scale radius can be constrained with a radial velocity for the stream with $r_{\rm apo} = 60$ kpc, though not as strongly as for the fiducial case. 
Despite the broadness of the posterior $r_s$ distribution for the furthest stream, the enclosed mass measurement $M_{\rm DM}[R = 60 \rm kpc] = 31.5^{+11.2}_{-10.7} \times 10^{10} \rm M_{\odot}$ is quite accurate and precise. Much like for the fiducial case, it also does not seem to be affected by the bias induced by a $r_s$ constraint.

When the stream track uncertainties are smaller, $r_s$ can be well recovered at every explored stream apocentric distance with the constraint increasing in accuracy and precision as the apocenter increases.  Consequently, the more precise $r_s$ constraints from streams with $r_{\rm apo} > 16$ kpc produce strongly biased $M_{\rm DM}$ posteriors.  Adding a radial velocity slightly improves the $r_s$ measurements, but does not significantly improve the accuracy for any scale radius or dark matter mass measurements.

\begin{figure}
    \centering
    \includegraphics[width=\columnwidth]{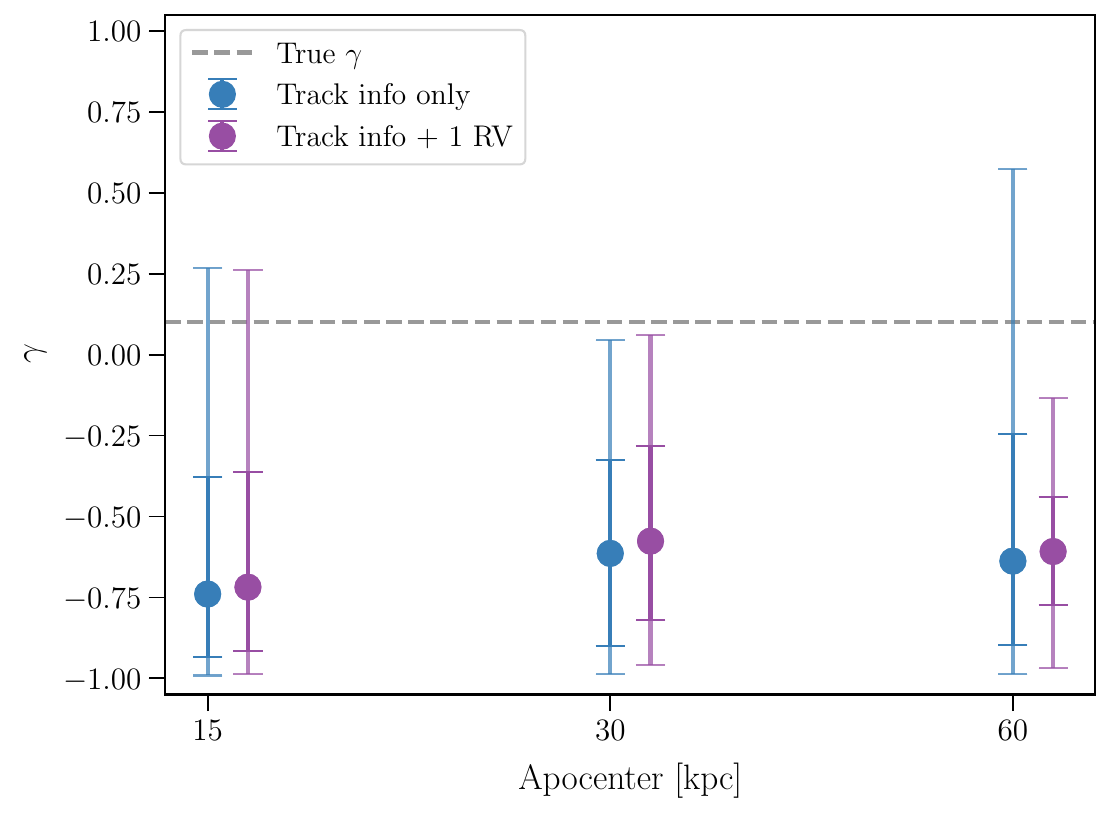}
    \caption{The dependence of radial profile constraints on the distance of the apocenter from the host galaxy of an observed stream in $\Phi_{\rm PL}$.  The precision of $\gamma$ improves as distance increases for fits with an RV, but no constraints can be made for any of the cases.}
    \label{fig:gamma_summary_varydist}
\end{figure}

Lastly, we find no correlation between the separation of the stream and host galaxy for constraining $\gamma$ in $\Phi_{\rm PL}$ as shown in Figure \ref{fig:gamma_summary_varydist}.  No constraints can be made as every case is dominated by degeneracies, however, we find the precision of the $\gamma$ measurement improves as the stream's apocenter gets farther away from the host when a radial velocity is added to the fit.       

\subsection{Changing Stream Eccentricity}
\label{eccresults}

\begin{figure*}
    \centering
    \includegraphics[width=0.8\textwidth] {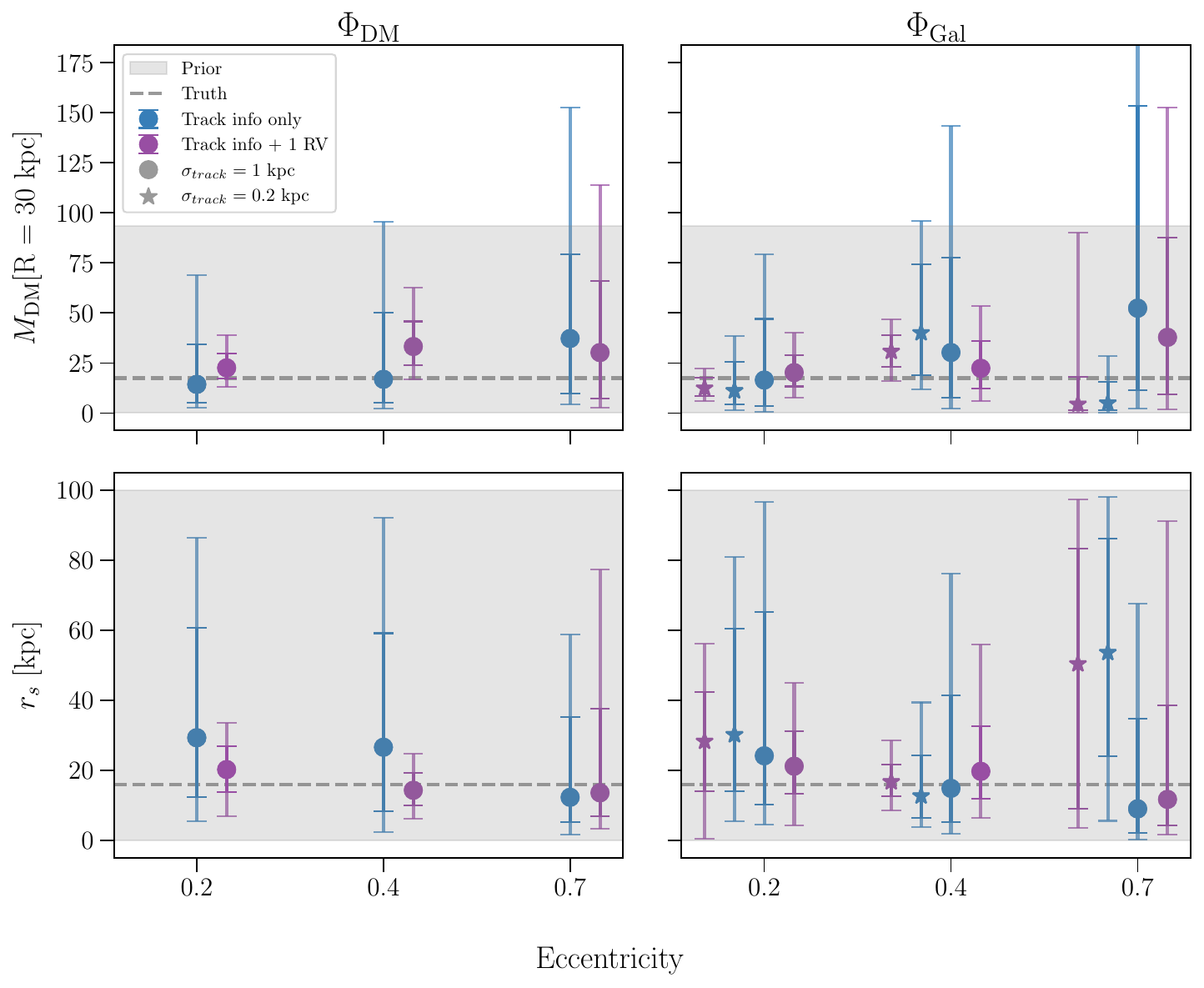}
    \caption{Same as Figure \ref{fig:mass_and_rs_summary_varylength} for varying stream eccentricities.}
    \label{fig:mass_and_rs_summary_varyecc}
\end{figure*}

We gauge the amount of information stored in a stream's eccentricity by varying the fiducial case to be more circular and more eccentric (last column Table \ref{tab:stream_props}).  Similar to those in Sections \ref{lengthresults} and \ref{incresults}, Figure \ref{fig:mass_and_rs_summary_varyecc} displays our findings for $r_s$ and $M_{\rm DM}$ for an NFW halo for the streams with varying eccentricities.  In $\Phi_{\rm DM}$, the only constraint to be made without a radial velocity is that $r_s$ is weakly constrained for the $e = 0.7$ stream case.  However, when a radial velocity is added, the scale radius is more strongly constrained for the fiducial and less eccentric streams.  Additionally, $M_{\rm DM}$ is strongly recovered for the $e = 0.2$ stream with only a slight bias from the $r_s$ constraint.

In $\Phi_{\rm Gal}$, no constraints can be made for stream data with $\sigma_{\rm track} = 1$ kpc until a radial velocity is added.  In which case, $r_s$ is then strongly recovered for the $e = 0.2$ stream.  Correspondingly, the enclosed DM mass measurement for this case, $M_{\rm DM}[R = 30 \rm kpc] = 20.2^{+8.8}_{-6.9} \times 10^{10} \rm M_{\odot}$, is incredibly accurate and precise. We find it is also the strongest mass constraint we with little bias from the $r_s$ constraint, which we will explain further in Section \ref{discussion}.      

With improved uncertainties, $r_s$ is weakly constrained only for the $e = 0.2$ stream without a radial velocity. Adding a radial velocity slightly tightens the previous $r_s$ constraints, but is not enough to help constrain $r_s$ for the most eccentric case.  A radial velocity measurement also allows the enclosed dark matter mass to be strongly constrained for the least eccentric case, again with little bias from the $r_s$ constraint.

\begin{figure}
    \centering
    \includegraphics[width=\columnwidth]{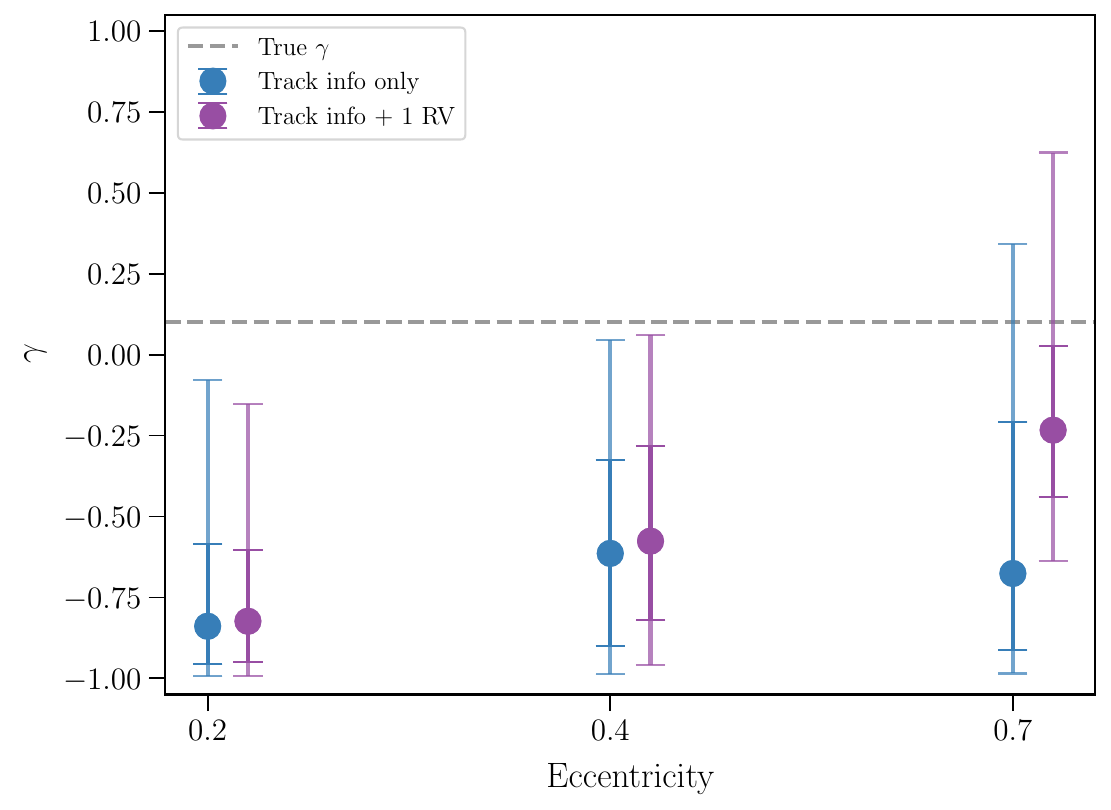}
    \caption{The dependence of radial profile constraints on the eccentricity of an observed stream in $\Phi_{\rm PL}$.  Measuring the radial profile improves as the eccentricity increases.}
    \label{fig:gamma_summary_varyecc}
\end{figure}

We display the constraints on $\gamma$ for these streams in $\Phi_{\rm PL}$ in Figure \ref{fig:gamma_summary_varyecc}, in which we find no constraints can be made except for the most eccentric case ($e = 0.7$) when a radial velocity included in the fit.  The constraint itself is weak, but the general trend of $\gamma$ as a function of eccentricity implies higher eccentricity orbits are better for recovering the radial profile of the potential.
\vspace{5pt}

We highlight our key results in Table \ref{tab:results_summary} in which we denote the best-observed stream cases where a constraint can be made on a host potential parameter.

\section{Discussion} \label{discussion}
We have shown that we can use extragalactic stellar streams to infer dark matter halo properties of galaxies outside of the Milky Way using only the `on-sky' positions of a stream track as well as with a single radial velocity measurement along the stream.  In this section, we address and explore the limitations of our approach (Section \ref{disc_degens_and_bias}), followed by what constraints can be made using only stream track measurements (Section \ref{disc_infocontent_streamtrackonly}), how much radial velocities help constraints (Section \ref{disc_infocontent_rvs}), as well as which stream properties are most informative (Section \ref{disc_infocontent_streamprops}).

\begin{figure*}
    \centering
    \includegraphics[width=\textwidth]{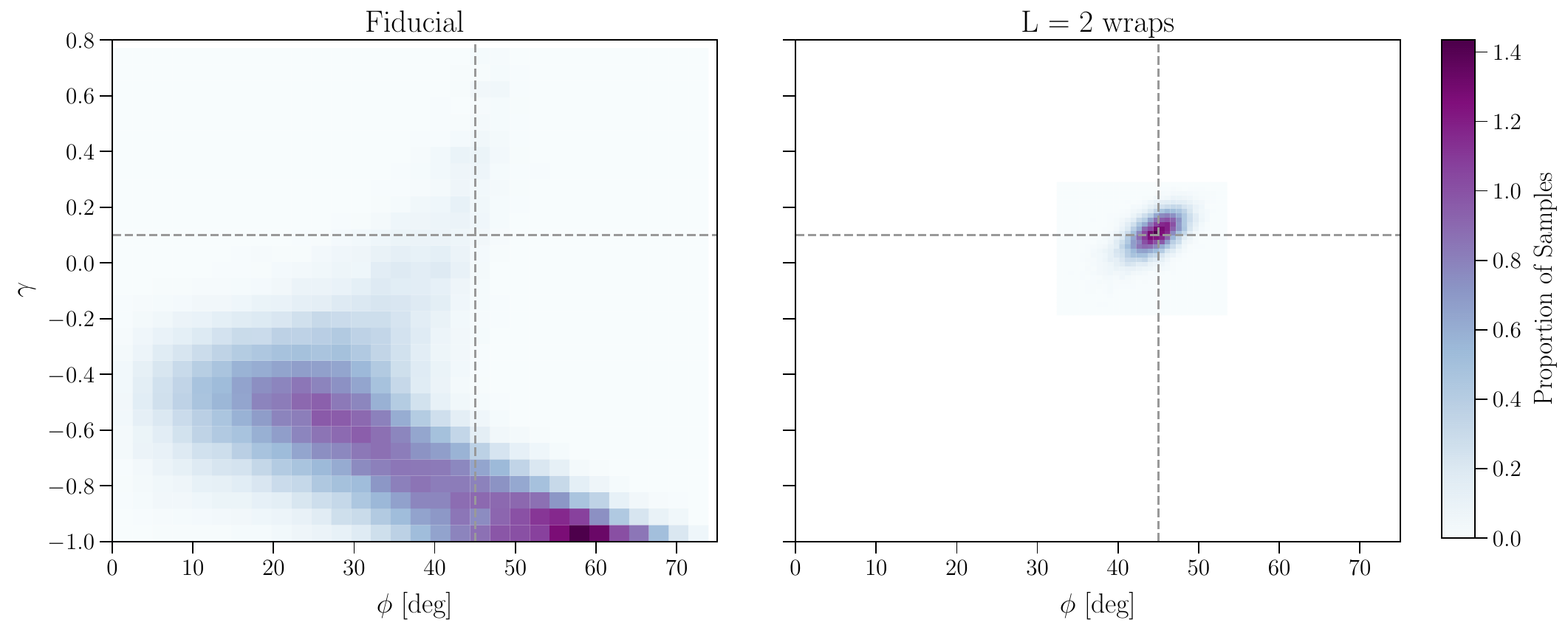}
    \caption{\textit{Left:} A 2-dimensional histogram of $\gamma$ samples versus the inclination angle $\phi$ of the fiducial stream fit without a radial velocity.  \textit{Right:} The same histogram of $\gamma$ samples as a function of inclination angle for the stream with 2 wraps and no radial velocity.  The true values of $\gamma$ and $\phi$ are denoted by dashed gray lines, while the darker colors correspond to a higher volume of samples, visualizing the trade-off between $\gamma$ and $\phi$ in the case where the stream does not have multiple loops and how this degeneracy is broken when the stream is long enough.}
    \label{fig:gamma_vs_incangle}
\end{figure*}

\subsection{Degeneracies and Biases}
\label{disc_degens_and_bias}

\subsubsection{Halo mass and progenitor velocity degeneracy}
\label{disc_mass_and_vel_degen}

As previously shown in Figure \ref{fig:cornerplot_PhiDM_example}, one degeneracy that we encounter in our fits is that between the dark matter halo mass and the initial velocity of the stream, where the mass increases indefinitely and scales roughly with the square of the velocity.  This is due to observing the streams in projection, meaning that multiple halo mass and stream velocity combinations can produce the same stream as observed in the sky.  Therefore, we limited the maximum halo mass to $200 \times 10^{10} M_{\odot}$ in our prior to better observe the behavior of the posterior in the regime of more realistic masses.

This degeneracy also biases the $r_s$ samples to lower values when fitting the stream with 2 wraps in $\Phi_{Gal}$.  The initial velocity of the progenitor is higher than it should be due to the model trying to compensate for the effect of the baryons on the orbit.  The model assumes the effects of the baryonic matter on the stream are negligible and thus tries to make a more massive and compact dark matter halo to account for these higher velocities.

\subsubsection{Halo mass and scale radius bias}
\label{disc_mass_and_rs_bias}

Many of our enclosed mass measurements for the streams in $\Phi_{\rm DM}$ and $\Phi_{\rm Gal}$ are biased to higher values in the cases where we are able to recover the scale radius with tight precision.  We investigate this by qualitatively comparing the prior and posterior distributions of the enclosed mass $M_{\rm DM}$ along with a distribution created from the mixture of the $r_s$ posterior and $M_{\rm NFW}$ prior for each stream case to determine whether the $r_s$ constraint is informing the $M_{\rm DM}$ posterior.  We find that for the cases with tight $r_s$ constraints and inflated $M_{\rm DM}$ measurements, the $M_{\rm DM}$ posterior is similar to that of the mixed distribution.  In general, we find that $M_{\rm DM}$ posteriors are more inflated for cases with stronger $r_s$ constraints, especially in the setups that are more informative, such as the streams with multiple wraps (Figure \ref{fig:mass_hist_2wraps}) and higher inclination angles.  

Additionally, the enclosed mass distributions in these cases look very precise, but this is also due to the effect of the $r_s$ constraint on the mass posterior.  In many cases, adding a single radial velocity helps lower the bias, though not enough to overcome it.  However, there are a few exceptions in which, despite the strength of a $r_s$ constraint, the enclosed mass posterior is affected very little, if at all, by the bias.  We will elaborate on these further in Section \ref{disc_infocontent_rvs}.        

\subsubsection{Radial profile in $\Phi_{\rm PL}$}
\label{disc_phi_PL_degens}
Many of our $\gamma$ results in $\Phi_{\rm PL}$ are affected by the tendency of the posterior samples to pile up at $\gamma = -1$.  Initially, we believed this to be due to a degeneracy between the stream's inclination angle and the potential, visualized in the left panel of Figure \ref{fig:gamma_vs_incangle}, in which an orbit with a certain inclination in one potential will look the same as an orbit in a certain potential with a different inclination angle when observed in projection.  We believed allowing the orbit to loop around multiple times allowed the MCMC to hone in on the correct solution in $\phi$-$\gamma$ space (Figure \ref{fig:gamma_vs_incangle} right panel).  However, we realize this does not completely explain why the MCMC prefers a $\gamma$ value of $-1$.  While the correlation between $\gamma$ and inclination does contribute to the runaway to some extent, we believe it is mainly due to our priors.  When $\gamma = -1$, a wide range of velocities can create orbits within our uncertainties which look the same when observed in projection, thus leading to a large prior volume that dominates the posterior.  As we will show in Section \ref{disc_infocontent_rvs}, including radial velocity measurements helps remedy this tendency. 

\begin{figure*}
    \centering
    \includegraphics[width=0.8\textwidth]{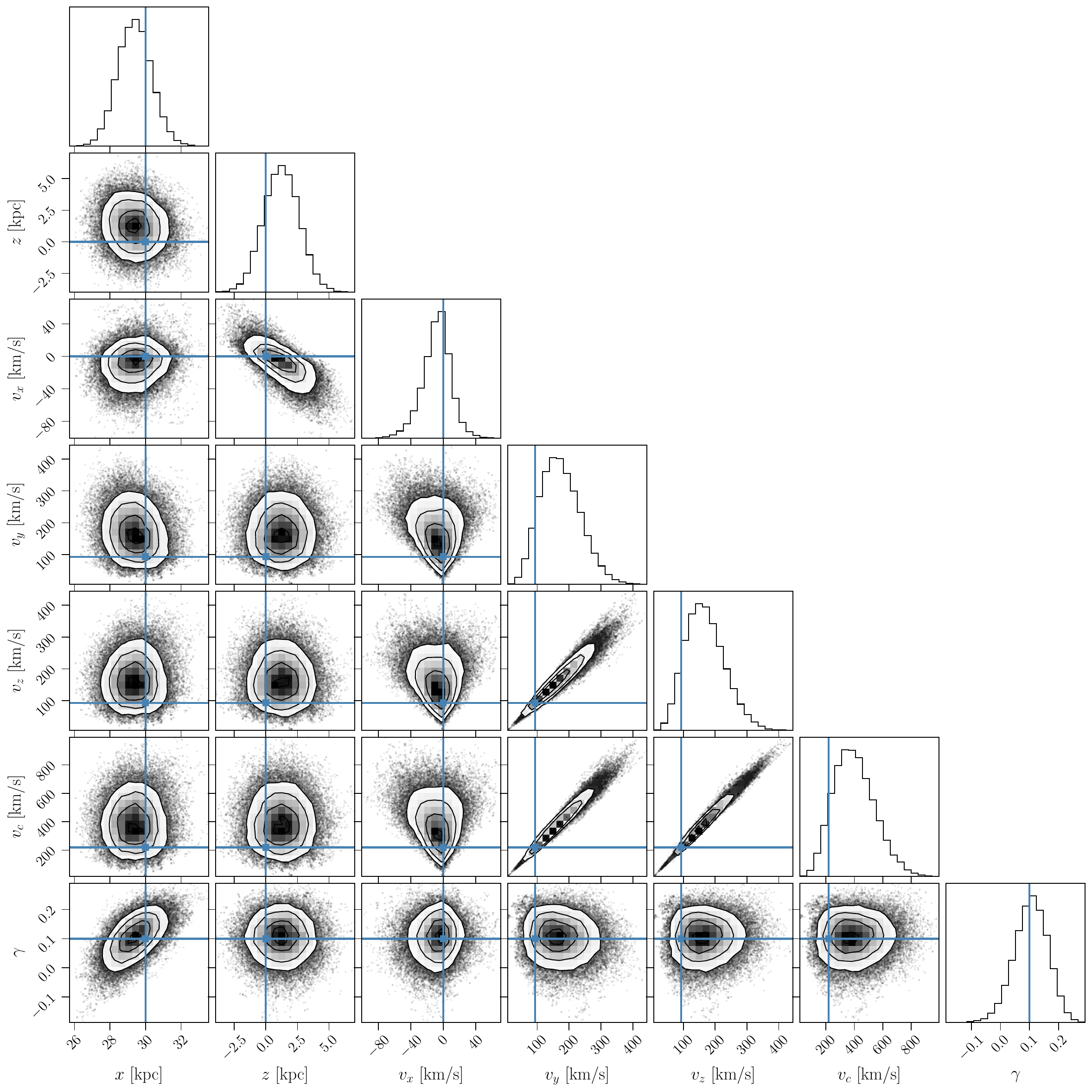}
    \caption{Corner plot of the MCMC samples for fitting the stream with two wraps parameters $x, z, v_x, v_y, v_z,$ and $\Phi_{PL}$ parameters $v_c, \gamma$ using only the mock observations of the stream track. The true values are indicated by the blue dots.}
    \label{fig:cornerplot_wraps}
\end{figure*}

\subsection{What can we learn with only the stream track?}
\label{disc_infocontent_streamtrackonly}

Despite the numerous degeneracies encountered with stream data limited only to `on-sky' positions, we have shown that it is still possible to infer properties of an extragalactic dark matter halo.  With current stream track uncertainties, the overall radial profile can be constrained as shown in Figure \ref{fig:cornerplot_wraps}, and a lower bound can \textit{tentatively} be placed on the enclosed dark matter mass when the stream has multiple well-measured wraps around its host galaxy.  While we acknowledge the biased nature of our measurement of $M_{\rm DM}$ in this case (discussed in sections \ref{disc_mass_and_vel_degen} and \ref{disc_mass_and_rs_bias}), we would like to note that the true enclosed mass is still within our 2$\sigma$ uncertainties.  Therefore, while it may not be a true measurement of the halo mass itself, it most certainly restricts the range of possible masses.  In Figure \ref{fig:dm_only_vs_gal_2_wraps_mass_hist}, we show the enclosed mass within 15 kpc distribution for this case in both $\rm \Phi{DM}$ and $\rm \Phi{Gal}$, in which the $M_{\rm DM}$ distribution in $\rm \Phi{Gal}$ has a clear boundary due to the baryonic component ($\sim \rm 7 \times 10^{10} M_{\odot}$), ruling out lower masses. 

In addition, the radial profile can be recovered for many stream cases with smaller stream track uncertainties, as well as the enclosed dark matter mass for a few, without needing a radial velocity measurement of the stream.

We note that previous studies have also shown that the flattening of a dark matter halo can also be inferred with stellar streams \citep{2023arXiv230317406N}. In that work, the radial profile of the external galaxy was fixed and only the flattening was varied. In contrast, in this work we have fixed the dark matter haloes to be spherical and only varied the radial profile. In future work, we will explore how well the radial profile and flattening can be simultaneously inferred using only the stream track. 

\subsection{How much do radial velocities help?}
\label{disc_infocontent_rvs}

\subsubsection{What can be done with a single radial velocity measurement?}
Adding just one radial velocity measurement to the fit often provides a significant boost to a streams constraining power on the host's dark matter halo.  In some cases, it adds power to make up for the lack of information from the intrinsic properties of a stream, i.e. half-wrap, low inclination, and low eccentricity stream cases.  In others, it strengthens an already existing measurement where there is enough information encoded in the stream, i.e. stream with 2 wraps in NFW haloes, as well as when stream track uncertainties are smaller.  

As previously suggested in \cite{Pearson_2022}, our results also imply just one radial velocity is needed to break degeneracies between halo parameters like halo mass and progenitor velocity.  We also emphasize that regardless of the stream's intrinsic properties, or improvement of observed stream track uncertainties, the halo mass cannot be genuinely measured without a radial velocity measurement.  While there are cases that seem to suggest the enclosed dark matter mass can be recovered precisely without a radial velocity, these are actually due to the strength of the constraint on $r_s$ which shrinks the range of possible halo masses.

In the cases where the enclosed dark matter mass is biased by the constraint on $r_s$, adding a radial velocity to the fit nearly always helps bring the measurement closer to the true mass, but is often not enough to break the bias.  Despite this, there are a few cases in which the enclosed dark matter mass seems to be genuinely recovered with a radial velocity.  The strongest of these cases being the low eccentricity (see Figure \ref{fig:mass_hist_eccpt2}) and fiducial streams in $\Phi_{\rm Gal}$, in which the $M_{\rm DM}$ posteriors are the most distinct from the enclosed mass distribution informed by the $M_{\rm NFW}$ prior and $r_s$ posterior.  These cases have the most accurate $M_{\rm DM}$ constraints with corresponding moderate $r_s$ constraints attainable with current observational uncertainties in $\Phi_{\rm Gal}$, making them the least affected by the aforementioned bias in a realistic galactic potential.  Other cases including a radial velocity in which $M_{\rm DM}$ is recovered with little bias from a $r_s$ constraint include: the $r_{apo} = 60$ kpc stream with $\sigma_{\rm track} = 1$ kpc, the $L = 0.5$ wraps, and the $\phi = 0^{\circ}$ streams with $\sigma_{\rm track} = 0.2$ kpc in $\Phi_{\rm Gal}$, and the $\phi = 70^{\circ}$ stream in $\Phi_{\rm DM}$.

For the streams in $\Phi_{PL}$ without 2 wraps, one radial velocity helps push the $\gamma$ posterior towards to the true value, however, it is often not enough to overcome the tendency for $\gamma$ to go to $-1$.  The exceptions for this are the $\phi = 0^{\circ}$ case in which the radial velocity helps overcome the degeneracy and $\gamma$ is constrained, and the $e = 0.7$ case in which the radial velocity boosts the $\gamma$ distribution high enough to weakly constrain it.  For the stream with $L = 2$ wraps, adding a single radial velocity measurement does not further improve upon the $\gamma$ constraint, but it does break the velocity degeneracies which allows the recovery of the progenitor's initial velocity and the circular velocity $v_c$ (see Figure \ref{fig:velocity_cornerplot_wraps}). 

\subsubsection{What happens as more radial velocity measurements are added?}
In addition to the data availability scenarios we test for our set of extragalactic streams, we also explore the impact of including more radial velocity measurements on our fits to the dark matter halo.  We do so by fitting the fiducial stream in each potential once again with $\sigma_{\rm track} = 1 \,{\rm kpc}$, and including 2 and then 3 radial velocity measurements.  Each additional radial velocity is taken at $\theta = \pm \frac{\pi}{2}$, respectively, and incorporated into each fit as an additional term to the likelihood as described in Section \ref{mock_obs_fit}.

\begin{figure*}
    \centering
    \includegraphics[width=0.8\textwidth] {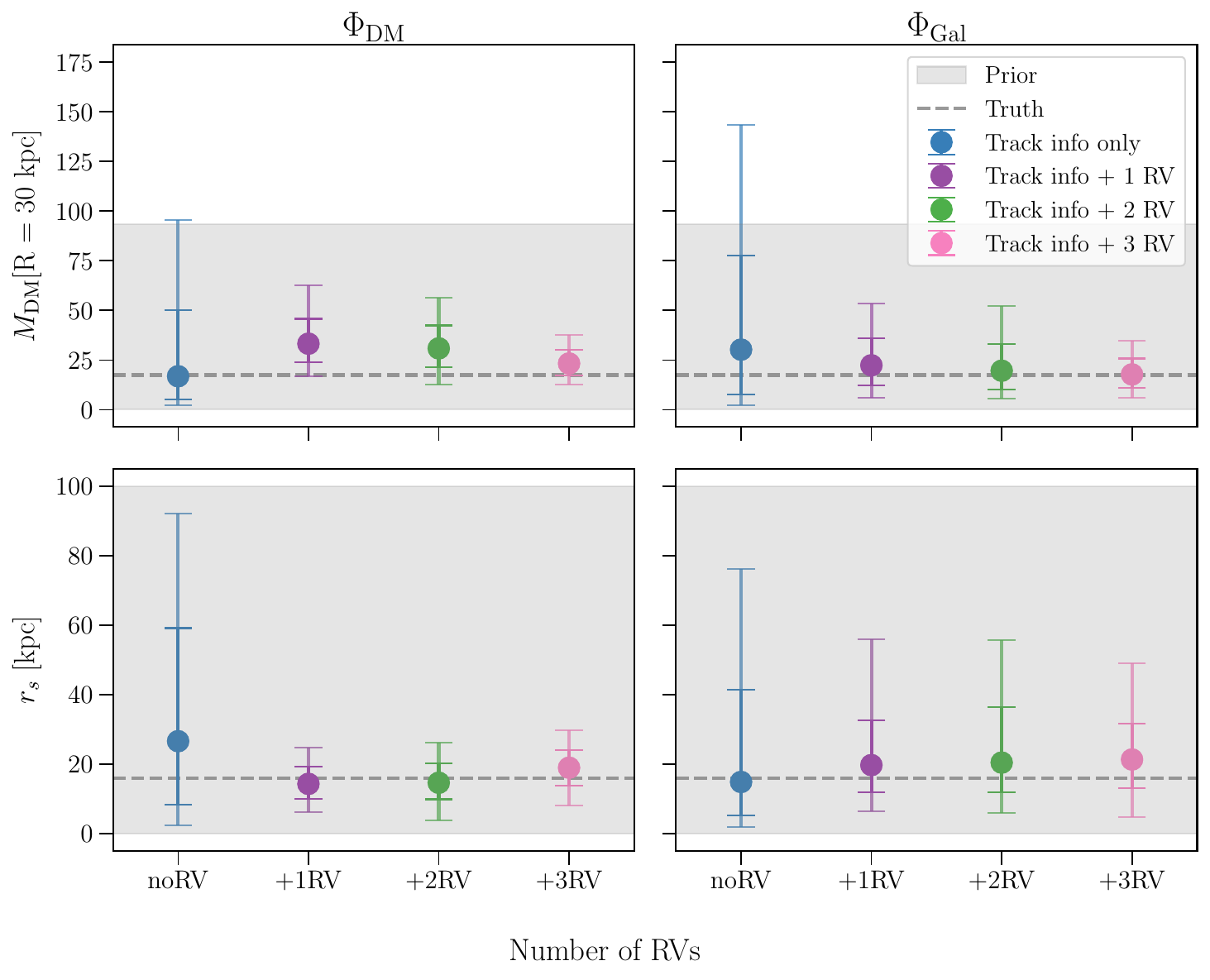}
    \caption{Similar in style to Figure \ref{fig:mass_and_rs_summary_varylength}, visualises the dependence of how well $M_{\rm DM}$ and $r_s$ are constrained as a function of the number of radial velocity measurements included when fitting the fiducial stream in $\Phi_{\rm DM}$ (left side) and $\Phi_{\rm Gal}$ (right side).}
    \label{fig:mass_and_rs_summary_morervs}
\end{figure*}

Figure \ref{fig:mass_and_rs_summary_morervs} shows our constraints on the enclosed mass and scale radius in $\Phi_{\rm DM}$ and $\Phi_{\rm Gal}$ for the fiducial stream with multiple RVs compared to the cases with just one or no radial velocities.  In $\Phi_{\rm DM}$, we find a second radial velocity only slightly improves upon the constraints made for $M_{\rm DM}$ and $r_s$ with one radial velocity, while the fit with a third radial velocity significantly improves the enclosed dark matter mass measurement in both accuracy and precision to be $M_{\rm DM}[R = 30 \rm kpc] = 23.3^{+6.8}_{-6.0} \times 10^{10} \rm M_{\odot}$, thus lessening the bias between $M_{\rm DM}$ and $r_s$.  Similarly, there is not much of a difference between the constraints made with one or two radial velocities in $\Phi_{\rm Gal}$, with a slight improvement in the enclosed dark matter mass measurement when the fit includes two radial velocities.  With a third radial velocity, there is a slight improvement in the precision of the $r_s$ measurement and an even slightly more improved mass measurement of $M_{\rm DM}[R = 30 \rm kpc] = 17.8^{+8.0}_{-6.6} \times 10^{10} \rm M_{\odot}$.  As was the case in $\Phi_{\rm DM}$, the additional radial velocity measurements aid in overcoming the mass and velocity degeneracy as well as the bias induced by a $r_s$ constraint.

\begin{figure}
    \centering
    \includegraphics[width=\columnwidth]{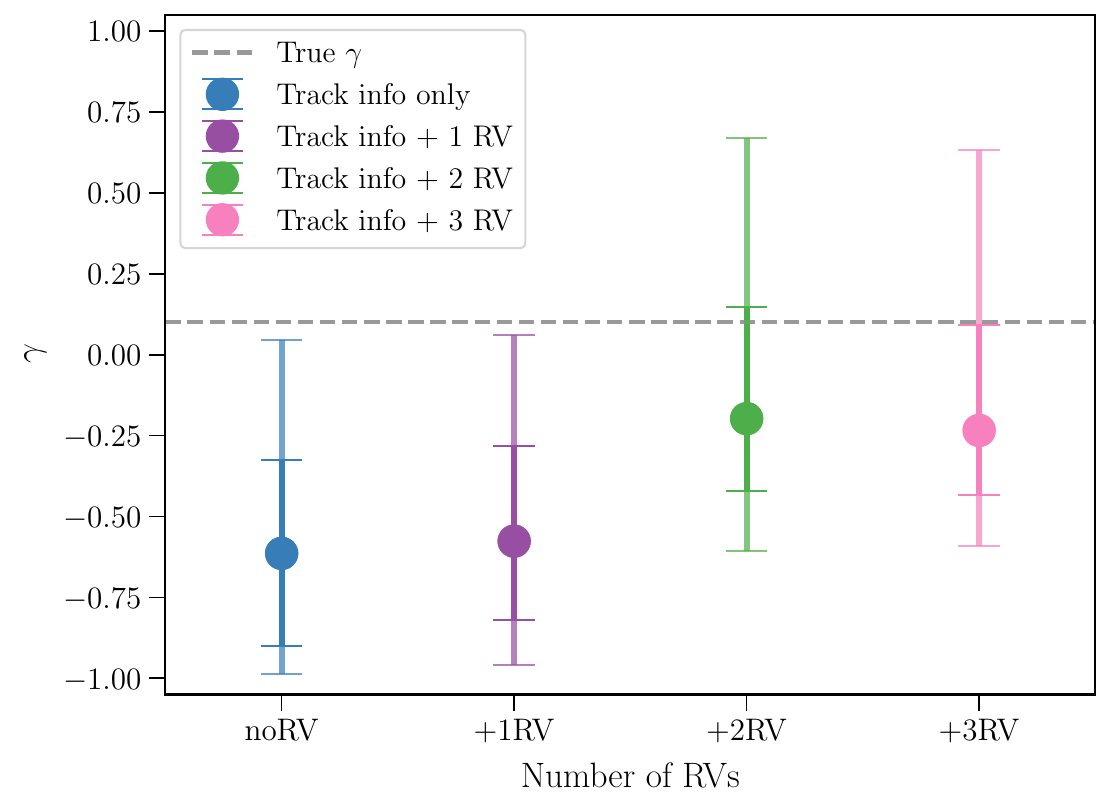}
    \caption{The dependence of how well $\gamma$ is constrained as a function of the number of radial velocity measurements included when fitting the fiducial stream in $\Phi_{\rm PL}$.  As before, the blue point corresponds to the fit without any radial velocities included, the purple point includes one radial velocity in the fit, green includes two, and pink includes three.}
    \label{fig:gamma_summary_morerv}
\end{figure}

Figure \ref{fig:gamma_summary_morerv} displays our results for these fits in $\Phi_{\rm PL}$. We find that a second radial velocity boosts the accuracy of the radial profile measurement in $\Phi_{\rm PL}$ enough to constrain it to $\gamma = -0.2^{+0.4}_{-0.2}$ with a similar precision as the fit with one radial velocity.  However, we also find the inclusion of a third radial velocity measurement provides about the same amount of constraining power on $\gamma$ as the fit with 2 RVs.  This indicates a non-linear relationship between the improvement of a constraint and the number of radial velocities included in the fit.

Overall, the inclusion of multiple radial velocity measurements in the fits serves to boost the strength of the constraints as well as helping to overcome degeneracies and biases.  However, the amount by which an additional radial velocity improves our measurements varies possibly due to the potential.  We find that in $\Phi_{\rm PL}$, an additional radial velocity significantly improves the accuracy of the constraint on $\gamma$.  While in $\Phi_{\rm DM}$ and $\Phi_{\rm Gal}$, additional radial velocities provide slight improvements upon the constraints made with just one radial velocity.  We would also like to note that the amount of help given by additional RVs could also vary based on the dark matter halo information gleaned from the intrinsic properties of the stream.  

Currently, we have to rely on the observability of a visible tracer, such as a planetary nebula or globular cluster, within an extragalactic stream to obtain even one radial velocity measurement.  Nonetheless, the upcoming spectrograph MOSAIC on the Extremely Large Telescope (ELT) is expected to be powerful enough to get spectra of individual extragalactic stream stars.  Its High Multiplex Mode will be sensitive to the Ca II infrared triplet lines for stream stars around at least 10 galaxies within 10 Mpc, ({Mart{\'\i}nez-Delgado} et al. 2023, in prep).  Therefore, it is most certainly probable that we will be able to obtain more than one radial velocity measurement for an extragalactic stream observed with MOSAIC.                 

\subsection{The most informative streams}
\label{disc_infocontent_streamprops}
In general, we find that longer streams, i.e. streams with more than one wrap around its host galaxy, provide the most accurate and precise constraints on the overall radial profile with and without the inclusion of a radial velocity measurement. When varying the inclination angle of the observed stream, we find streams with lower inclinations better constrained $\gamma$ when modelled in $\Phi_{\rm PL}$, while streams with higher inclinations yield better $r_s$ constraints when modelled in $\Phi_{\rm DM}$ and $\Phi_{\rm Gal}$. When varying stream eccentricity, we find that streams on more eccentric orbits $(e > 0.4)$ provide better constraints on the radial profile when modelled in $\Phi_{\rm PL}$, while streams on moderate to less eccentric orbits provide better constraints on $M_{\rm DM}$ and $r_s$ in $\Phi_{\rm DM}$ and $\Phi_{\rm Gal}$.  Finally, we find that streams with apocentres beyond the scale radius of the dark matter halo are better for radial profile and enclosed DM mass constraints.

\section{Conclusions}

In this work, we have explored how well the properties of dark matter haloes can be measured using observations of extragalactic streams with various properties in the regime of limited data availability (2D and 3D). By generating and fitting a set of possible mock streams around different potentials with observations of varying data availability, we have demonstrated that it is possible to infer extragalactic dark matter halo properties with stellar streams, despite the lack of high-quality data.\\

\noindent Using only the stream's position on the sky:

\begin{itemize}
    \item We find that with stream track uncertainties possible with current observing facilities, the radial profile of the dark matter potential can be precisely constrained \textbf{without} any radial velocity measurements when a stream has multiple measured wraps around the host galaxy.
    \item For the same stream case, we find tentative evidence that the enclosed mass may be bounded, but not truly measured.  This is due to the precision and accuracy of the radial profile constraint influencing the enclosed mass constraint in some cases, rather than the constraint fully being due to the measurement of the enclosed mass
    \item When assuming stream track uncertainties expected to be possible with future photometry, we find that the radial profile can be recovered for many observable stream scenarios.
\end{itemize}

\noindent With just one radial velocity measurement along the stream:
\begin{itemize}
    \item We find that the dark matter mass enclosed within the distance to the stream from the host can be accurately measured
    \item We also find a boost to constraining power for all explored DM halo properties with many streams, aid in overcoming degeneracies and biases, and strengthened existing constraints. 
\end{itemize}

\noindent We also find that different stream properties store different amounts of information depending on the dark matter potential parameter you are trying to measure as well as the assumed form of the dark matter potential.  Generally:
\begin{itemize}
    \item Longer streams, i.e. streams with multiple measurable wraps around their host, provide the most information on the overall radial profile, regardless of its form.
    \item Following longer streams, we find that the most information on the radial profile of a DM potential is determined by a stream's inclination angle, followed by its eccentricity, and then apocenter.
    \item When aiming to measure the dark matter mass enclosed, streams with low eccentricities, followed by those with moderate orbital properties, then those with apocenters in the outer halo are the most informative.
\end{itemize}

\noindent Finally, additional radial velocity measurements further strengthen constraints and aid in overcoming degeneracies and biases.

In the future, we plan to employ our approach to streams in cosmological simulations, with observations reflective of what we expect from future surveys/telescope missions, such as LSST, NGRST, and ARRAKIHS.  Additionally, in pursuing our goal of applying our stream-fitting techniques to real extragalactic stream data, we will apply our method to that observed in current surveys, such as the aforementioned Stellar Stream Legacy Survey, which will give us the opportunity to probe dark matter haloes of with nearly 100 external streams for the first time.

%The last numbered section should briefly %summarise what has been done, and describe
%the final conclusions which the authors %draw from their work.

\section*{Acknowledgements}

DE acknowledges support through ARC DP210100855.

This research was done using the sampling tools from {\fontfamily{qcr}\selectfont zeus} \citep{2021MNRAS.508.3589K} and the additional {\fontfamily{qcr}\selectfont PYTHON} packages for analysis and visualization: {\fontfamily{qcr}\selectfont numpy} \citep{harris2020array}, {\fontfamily{qcr}\selectfont scipy} \citep{2020SciPy-NMeth}, {\fontfamily{qcr}\selectfont h5py} \citep{collette_python_hdf5_2014}, {\fontfamily{qcr}\selectfont matplotlib} \citep{Hunter:2007}, and {\fontfamily{qcr}\selectfont corner} \citep{corner}.

%The Acknowledgements section is not %numbered. Here you can thank helpful
%colleagues, acknowledge funding agencies, %telescopes and facilities used etc.
%Try to keep it short.

%%%%%%%%%%%%%%%%%%%%%%%%%%%%%%%%%%%%%%%%%%%%%%%%%%
\section*{Data Availability}

The fits in this work will be made available upon reasonable request to the corresponding author.

%The inclusion of a Data Availability Statement is a requirement for articles published in MNRAS. Data Availability Statements provide a standardised format for readers to understand the availability of data underlying the research results described in the article. The statement may refer to original data generated in the course of the study or to third-party data analysed in the article. The statement should describe and provide means of access, where possible, by linking to the data or providing the required accession numbers for the relevant databases or DOIs.

\begin{table*}
\caption{Summary of the best stream cases with current stream track uncertainties where a host potential parameter is measured, in order from strongest to weakest constraint.}
\label{tab:results_summary}
\begin{adjustbox}{width=\textwidth}
\begin{tabular}{|l|l|l|l|}
\hline
Parameter & Best Stream Cases (with $\sigma_{\rm track} = 1$ kpc) & Track info only & Track info + 1 radial velocity  \\ \hline
$\gamma$ & $L = 2$ wraps & Constrained & Constrained     \\ 
& low inclination & Unconstrained & Constrained \\
& high eccentricity & Unconstrained & Constrained \\
\hline
$r_{s}$  & $L = 2$ wraps & Constrained & Constrained \\ 
& high inclination & Constrained & Constrained\\
& low eccentricity & Unconstrained & Constrained \\
& fiducial & Unconstrained & Constrained \\
& $r_{\rm apo} = 60$ kpc & Unconstrained & Constrained \\
& $L = 0.5$ wrap & Unconstrained & Constrained \\
\hline
$M_{\rm NFW}$ & low eccentricity & Unconstrained & Constrained \\
& fiducial & Unconstrained & Constrained \\
& $r_{\rm apo} = 60$ kpc & Unconstrained & Constrained \\
\hline
\end{tabular}
\end{adjustbox}
\end{table*}

%%%%%%%%%%%%%%%%%%%% REFERENCES %%%%%%%%%%%%%%%%%%

% The best way to enter references is to use BibTeX:

\bibliographystyle{mnras}
\bibliography{references} % if your bibtex file is called example.bib

% Alternatively you could enter them by hand, like this:
% This method is tedious and prone to error if you have lots of references
%\begin{thebibliography}{99}
%\bibitem[\protect\citeauthoryear{Author}{2012}]{Author2012}
%Author A.~N., 2013, Journal of Improbable Astronomy, 1, 1
%\bibitem[\protect\citeauthoryear{Others}{2013}]{Others2013}
%Others S., 2012, Journal of Interesting Stuff, 17, 198
%\end{thebibliography}

%%%%%%%%%%%%%%%%%%%%%%%%%%%%%%%%%%%%%%%%%%%%%%%%%%

%%%%%%%%%%%%%%%%% APPENDICES %%%%%%%%%%%%%%%%%%%%%

\appendix

\section{Extra Plots}

\begin{figure}
    \centering
    \includegraphics[width=\columnwidth]{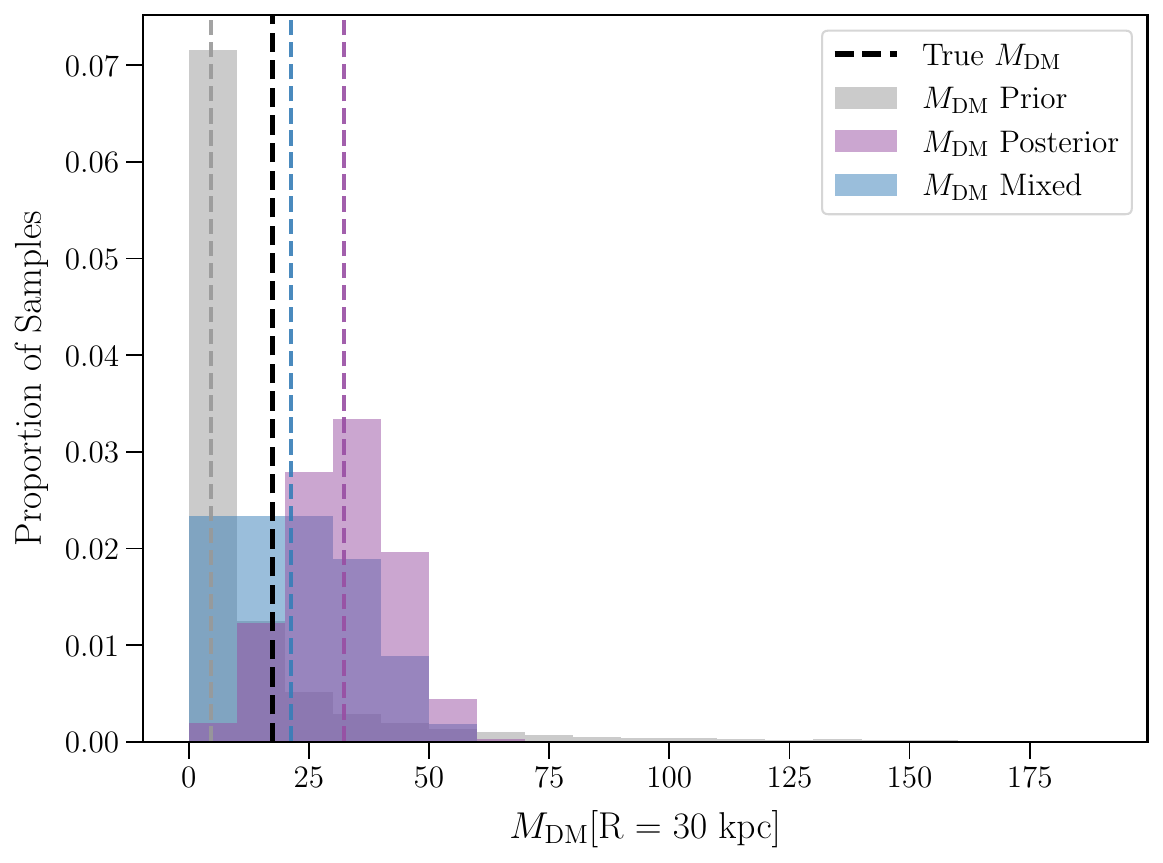}
    \caption{Normalized histograms displaying the enclosed mass prior (gray), posterior (purple), and mixed (prior informed by $r_s$ posterior in blue) distributions for the stream case with 2 wraps and no radial velocity in $\Phi_{\rm DM}$.  Each histogram has a dashed line of the corresponding color denoting the median of the distribution.  The true enclosed dark matter mass is denoted by the black dashed line.}
    \label{fig:mass_hist_2wraps}
\end{figure}

\begin{figure*}
    \centering
    \includegraphics[width=\textwidth]{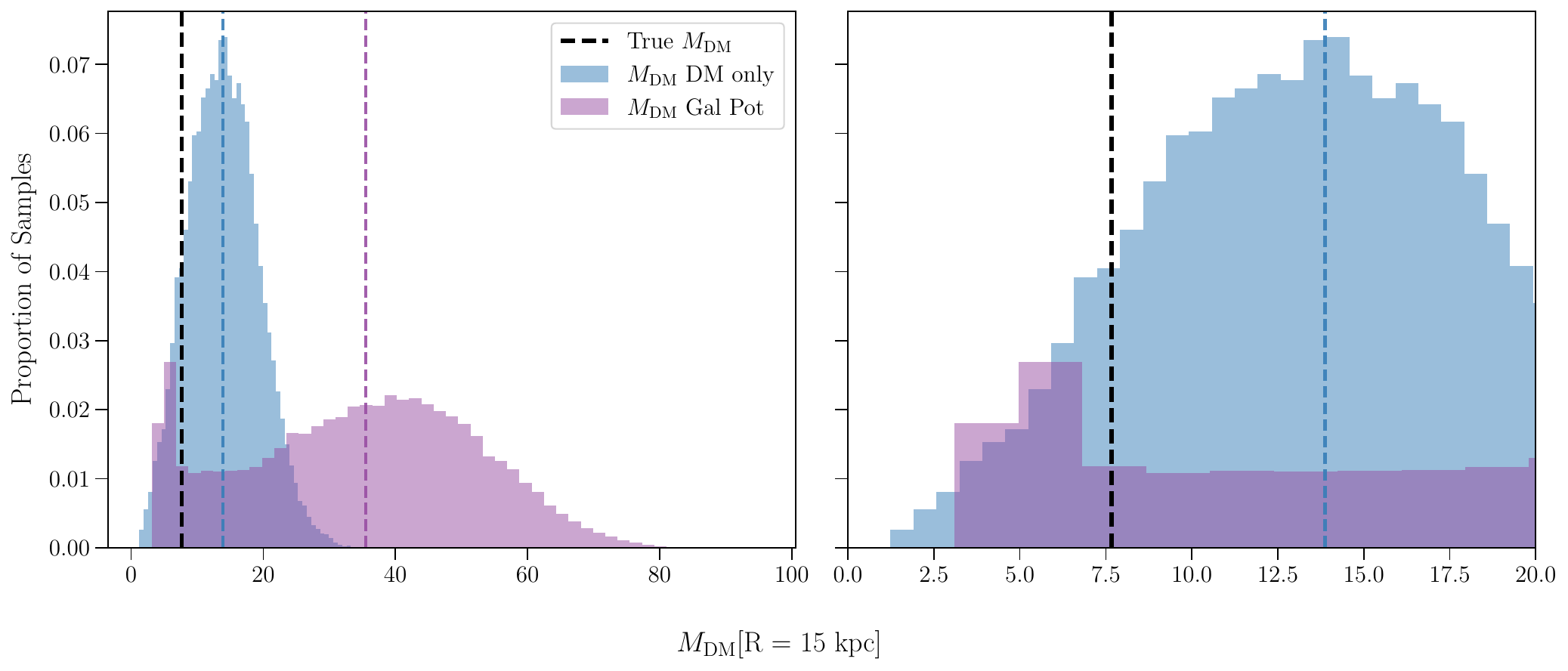}
    \caption{\textit{Left:} Normalized histograms displaying the enclosed dark matter mass posterior distributions within 15 kpcs for the stream case with 2 wraps and no radial velocity in $\Phi_{\rm DM}$ (blue) and $\Phi_{\rm Gal}$ (purple) with medians indicated by the dashed vertical lines of the corresponding color.  The truth is denoted by a dashed black line. \textit{Right:} a zoomed in version of the left panel showing the lower mass range ($ \leq 20 \times 10^{10}\rm M_{\odot}$) of both distributions.}
    \label{fig:dm_only_vs_gal_2_wraps_mass_hist}
\end{figure*}

\begin{figure}
    \centering
    \includegraphics[width=\columnwidth]{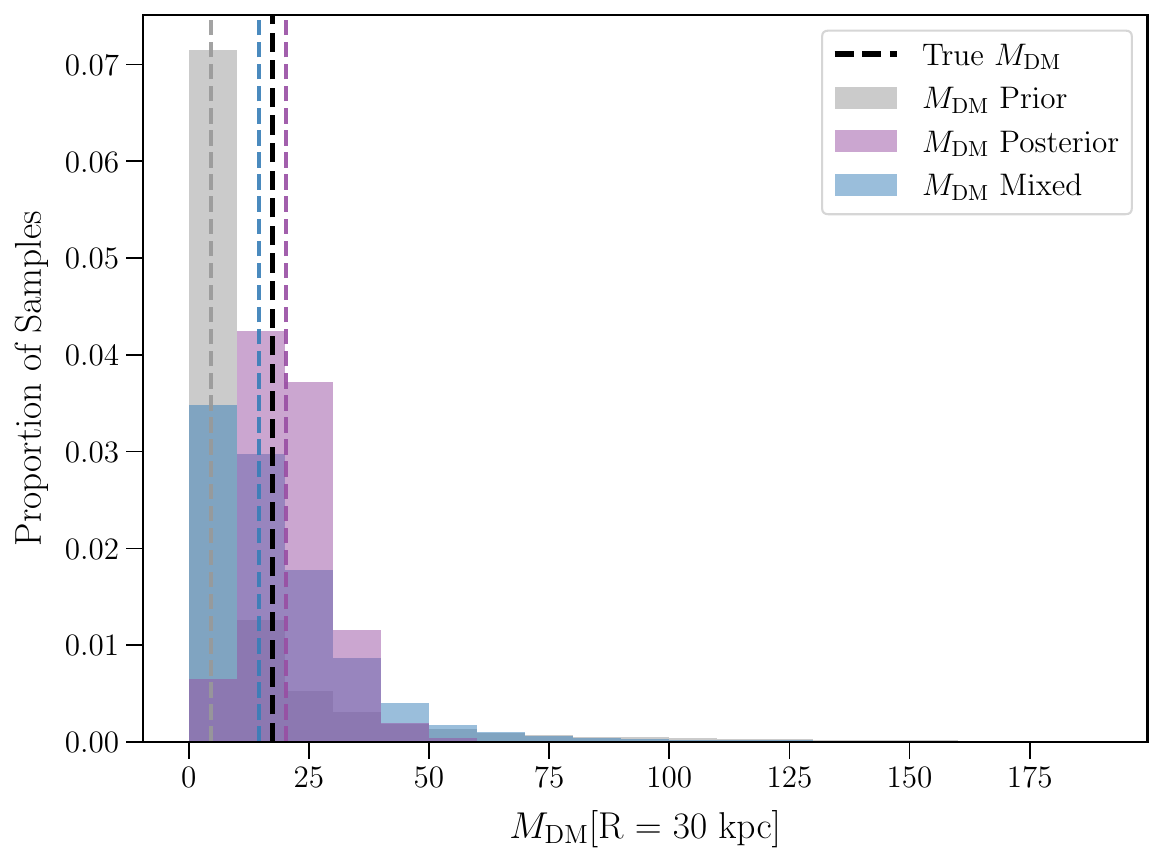}
    \caption{Normalized histograms displaying the enclosed mass prior, posterior, and mixed (prior informed by $r_s$ posterior) distributions for the $e = 0.2$ stream case with one radial velocity in $\Phi_{\rm Gal}$.  The colors and labels denote the same aspects of the plot as in Figure \ref{fig:mass_hist_2wraps}.  $M_{\rm DM}$ posterior in this case seems to be the least affected by the bias induced by a $r_s$ constraint.}
    \label{fig:mass_hist_eccpt2}
\end{figure}

\begin{figure}
    \centering
    \includegraphics[width=\columnwidth]{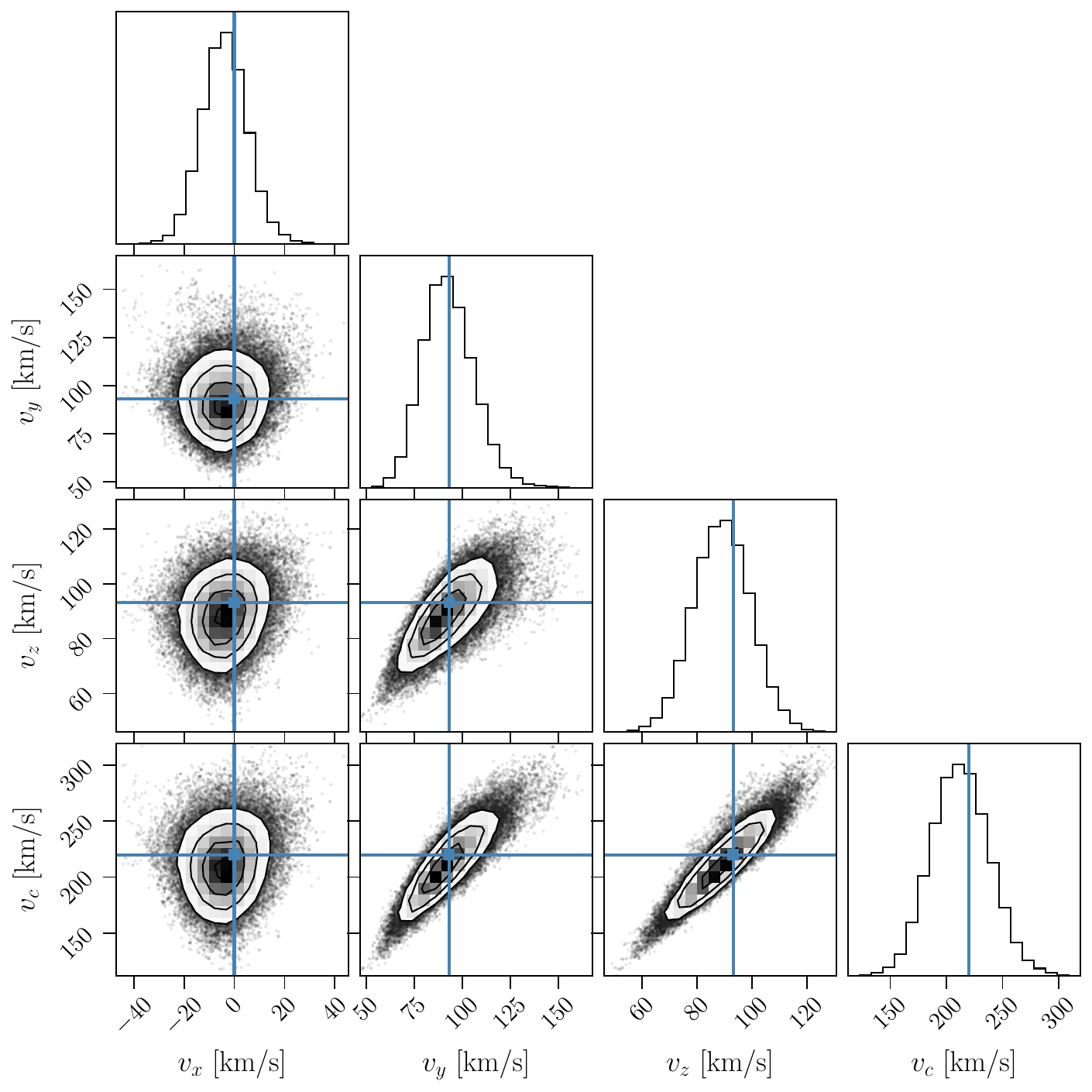}
    \caption{Corner plot of the velocity MCMC samples for fitting the stream with multiple wraps parameters $v_x, v_y, v_z,$ and $v_c$ in $\Phi_{PL}$ using mock observations of the stream track with a single radial velocity measurement.  Illustrates how the velocity degeneracies seen in Figure \ref{fig:cornerplot_wraps} are broken once a radial velocity is added to the fit.}
    \label{fig:velocity_cornerplot_wraps}
\end{figure}

%%%%%%%%%%%%%%%%%%%%%%%%%%%%%%%%%%%%%%%%%%%%%%%%%%

% Don't change these lines
\bsp	% typesetting comment
\label{lastpage}
\end{document}